# Emergent conservation in atmospheric chemical mechanisms

**Beatriz Lucia G. Rodriguez[1†], Patrick Obin Sturm[1†], Daniel Getter[1], and Sam J. Silva[1,2,3]**

[1]Department of Earth Sciences, University of Southern California, Los Angeles, CA, USA

[2]Department of Environmental Engineering, University of Southern California, Los Angeles, CA, USA

[3]Department of Population and Public Health Sciences, University of Southern California, Los Angeles, CA, USA

Corresponding author: Patrick Obin Sturm (psturm@usc.edu)

†Equally contributing authors

**Key Points:**

- Atmospheric chemical mechanisms contain emergent conserved quantities independent of mass conservation that reduce their dimensionality
- These relationships form a unified class of conserved quantities induced by coupled kinetics, which we term kinetic invariants
- Emergent kinetic invariants appear in 15 out of 35 surveyed mechanisms and reveal hidden relationships between species

**Abstract**

Conservation laws are time-invariant properties that constrain many physical systems. For systems of chemical reactions, conservation of mass constrains how atoms flow between chemical species. Here, we demonstrate that chemical reaction networks can display emergent conserved quantities not explained by mass conservation that arise instead from coupled kinetics. While some of these kinetic relationships can be immediately recognizable in simple chemical systems, more complex mechanisms have hidden nontrivial relationships between species. We identify these emergent relationships as a broader class of invariant behavior in atmospheric chemical mechanisms, which we call kinetic invariants. Kinetic invariants emerge when branching reactions with proportional rates cause species concentrations to evolve in lockstep, and can be found across atmospheric chemical mechanisms spanning orders of magnitude in complexity. Identifying kinetic invariants can provide theoretical bounds for exact mechanism reduction and reveal underlying relationships in the atmospheric chemical system.

**Plain Language Summary**

Complex systems like the chemistry of the atmosphere have underlying symmetry and simplicity, where even under rapidly changing conditions, some things stay the same. One example of such a constant in time is the physical law of mass conservation, where matter cannot be created or destroyed, only rearranged in chemical reactions. However, mass conservation is not the only time-constant behavior in chemistry: we sometimes find emergent conservation that can't be explained by conservation of mass. Emergent conservation appears when certain reactions depend on the same chemical species, causing the rates of change of some species to move in lockstep with each other. We can use this principle to reveal previously unknown dependencies and hidden relationships in atmospheric chemistry.

## 1 Introduction

The dynamic behavior of complex systems in the physical world is often constrained by time-invariant properties such as conservation laws. Conservation laws are statements of fundamental symmetry and simplicity within complex systems, and are represented mathematically as conserved quantities that remain constant over time regardless of how quickly the system itself changes. A canonical example is the law of mass conservation, attributed to the works of Mikhail Lomonosov and Antoine Lavoisier: matter is not created or destroyed in chemical reactions, only transformed as atoms are rearranged across chemical species (Lavoisier, 1789). In systems of chemical reactions, conservation of mass constrains the stoichiometry of reactions to preserve atom balances, which are well-known conserved quantities. In the dynamic system that governs the chemical composition of the atmosphere, matter flows through a complex network of reactive chemical species whose behavior and time evolution are of societal and environmental importance (Seinfeld & Pandis, 2016).

Concepts of invariance recur throughout atmospheric chemistry and in the numerical chemical mechanisms built to represent it, leading to additional relationships between chemical species that help constrain the complex system. The Leighton relationship relating ozone to nitrogen oxides under photostationary conditions is an example of a pseudo-steady state assumption that, while only approximately time-invariant, identifies near-equilibrium behavior to obtain algebraic relationships between species (Leighton, 1961). Chemical families such as odd-nitrogen ($NO_y$) or odd oxygen ($O_x$, or the expanded family $O_y$) group species that rapidly interconvert on timescales where their total is approximately conserved, providing tractable budgets for reactive nitrogen and ozone (Bates & Jacob, 2020; Seinfeld & Pandis, 2016). Improved diagnostics for ozone production in chemical mechanisms have been built based on conservation of electron spin (Edwards & Evans, 2017). Elemental mass conservation, perhaps the most fundamental time-constant behavior, constrains how atoms flow between species and manifests as a set of stoichiometric invariants: exactly conserved combinations of species concentrations that are weighted by atomic composition. All of these time-invariant constraints provide useful relationships that constrain the dynamics of the complex system. However, they alone do not fully characterize all time-invariant behavior in the system.

Here we show that atmospheric chemical mechanisms contain a distinct class of emergent conserved quantities that we term “kinetic invariants”. Kinetic invariants arise not from stoichiometry but from coupled kinetics, where branching reactions create dependencies between the rates of change of different species. The simplest case of this kind of invariant, when products have a fixed yield ratio, is familiar from chamber studies and can be used to characterize branching along different pathways (Aschmann & Atkinson, 1994; Nguyen et al., 2016). We show that this easily recognizable fixed ratio of products is a special case of a broader class of conserved quantities. The same core principle is also responsible for relationships that are not recognizable by inspection, one of which appeared in a data-driven analysis that found more invariants in a chemical mechanism than structural stoichiometry could explain (Liu et al., 2024) and was later linked to mass action kinetics (Blokhuis et al., 2025, 2026). We find that such relationships can be chemically rich, linking conserved groups of species across multiple generations of chemistry and seemingly independent pathways. This unified framework of kinetic invariants provides a common explanation for relationships across this full range of complexity, from a single branching reaction to higher-order connectivity in the reaction network. We use this framework to systematically survey atmospheric chemical mechanisms for emergent kinetically-defined families that are exactly conserved.

## 2 Identifying kinetic invariants: from a simple system to a general approach

The simplest kinetic invariants are pairwise relationships between products of a branching reaction. We isolate this case in a deliberately minimal but illustrative example of $RO_2$+NO branching reactions, important in the photochemical production of ozone and secondary aerosol (Browne et al., 2013). Nitric oxide NO oxidized by a peroxy radical $RO_2$ forms an alkoxy radical RO and nitrogen dioxide $NO_2$ to propagate ozone-forming cycles, or alternatively forms an organic nitrate $RONO_2$ with branching probability $\alpha_2$. RO reverts rapidly to $RO_2$ in the Earth's oxygen-rich atmosphere. This 5-species, 3-reaction system is

$$RO_2 + NO \rightarrow RO + NO_2, \qquad r_1 = k_1[RO_2][NO] \tag{1}$$
$$RO_2 + NO \rightarrow RONO_2, \qquad r_2 = k_2[RO_2][NO] \tag{2}$$
$$RO \overset{O_2}{\rightarrow} RO_2, \qquad r_3 = k_3[RO] \tag{3}$$

where reaction rates $r_1, r_2, r_3$ are proportional to reactant concentrations via rate constants $k_1, k_2, k_3$: this kinetic formulation of reaction rates is known as the mass action law (Guldberg & Waage, 1879).

Assuming that the system is closed (i.e., it is isolated from external sinks and sources) and reactions 1-3 are the only reactions that occur, there are three invariant quantities present. The first two are familiar invariants due to stoichiometry: conservation of atoms (i.e., N and the R moiety are conserved). The stoichiometric invariants are detectable by accounting for species atomic composition or investigating the static structure of atom-balanced reactions. The third constraint is not detectable from the static reaction network structure alone. However, the mechanism is not static, and the kinetics impose a third constraint arising because reaction rates $r_1$ and $r_2$ are proportional:

$$k_2 d_t[NO_2] - k_1 d_t[RONO_2] = 0 \tag{4}$$

This conserved quantity emerging from equations 1-3 is the simplest member of a class we call "kinetic invariants": conserved combinations of species rates that emerge from coupled rates, weighted by kinetic parameters that no elemental balance could produce. In a closed system with no other processes, the rate form of a kinetic invariant holds at every moment across all environmental conditions (e.g. temperature and in some cases pressure, actinic flux, and relative humidity). Even as reaction rates $r_1$ and $r_2$ change their proportionality, equation 4 remains a constraint on the dynamics, because environmental dependency is included in its weights $k_1(T)$ and $k_2(T)$. When environmental conditions and rate constants are fixed, as within a chemical operator timestep in a 3D model or a controlled chamber experiment, equation 4 integrates to a combination of concentrations $k_2[NO_2] - k_1[RONO_2]$ that stays constant in time. Kinetic and stoichiometric invariants act as independent constraints on the chemical system, limiting its accessible states to their intersection as visualized in Figure 1.

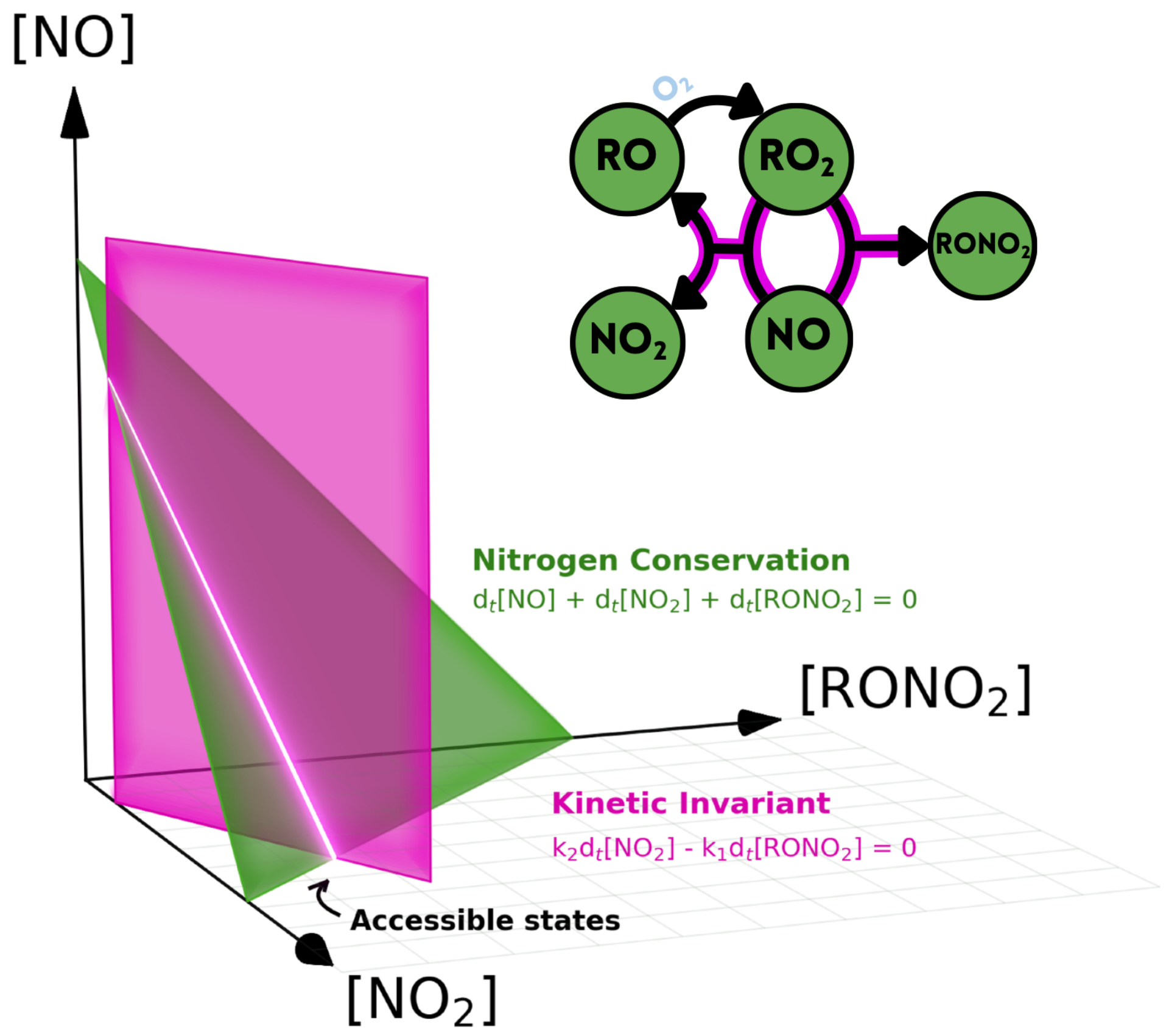

**Figure 1.** The time evolution of the simple organic nitrate system is constrained by conserved quantities: stoichiometric mass conservation (here the simplex of nitrogen species) and a kinetic invariant, a linearly independent surface relating $NO_2$ and organic nitrate through kinetic parameters. Together these constraints limit the accessible states in the system to their intersection, here projected onto species containing nitrogen.

Conserved quantities determine the intrinsic dimensionality of chemical systems. The time evolution of the organic nitrate system in equations 1-3 could be written as a system of five coupled ordinary differential equations (ODEs), seemingly with five degrees of freedom. However, due to stoichiometric and kinetic invariants, this system is constrained to evolve on a two-dimensional manifold within the five-dimensional space. In Text S1 we derive that latent space using only two ODEs that evolve the system forward in time identically to the five-equation system. With the invariants and initial concentrations, the concentrations of all five species can be reconstructed from two species at any time. This is the kind of low-dimensional latent structure that reduced-order and data-driven models of atmospheric chemistry are thought to leverage, and kinetic invariants strengthen the theoretical basis for this underlying simplicity (Bodnar et al., 2025; Kelp et al., 2020; Liu et al., 2024; Wiser et al., 2023, 2025; Yang et al., 2024, 2025).

Kinetic invariants can reduce the effective number of species; proportional reaction rates always reduce the effective number of reactions. As is commonly done when designing chemical reaction mechanisms, reactions 1 and 2 can equivalently be represented by a merged effective reaction:

$$RO_2 + NO \rightarrow \alpha_1 RO + \alpha_1 NO_2 + \alpha_2 RONO_2, \qquad r_m = \ (k_1 + k_2)[RO_2][NO] \tag{5}$$

with an overall reaction rate $r_m$, conserving mass with pseudo-stoichiometric weights $\alpha_1 = \frac{k_1}{k_1+k_2}$ and $\alpha_2 = 1 - \frac{k_1}{k_1+k_2} = \frac{k_2}{k_1+k_2}$ that are branching probabilities of the two reaction channels. It is important to note that the kinetic invariants described here are an intrinsic property of the chemistry. They exist regardless of the design choice to represent branching reactions as multiple elementary reactions or one equivalent merged reaction. When branching reactions are already merged with fixed branching ratios, a kinetic invariant may appear in the static structure of the mechanism as a structural invariant with non-integer coefficients even though its origin is kinetic, rather than an emergent invariant hidden to the structure. Regardless of mechanism choice, the necessary condition for kinetic invariants is that reactions are merge*able* because of proportional rates. Mergeability is necessary but not sufficient for kinetic invariants: proportional reactions create either species-level redundancy, leading to kinetic invariants, or are absorbed by existing reaction-level redundancy, removing an equilibrium null cycle (Blokhuis et al., 2025; Text S2). Which outcome depends on system-wide connectivity that is rarely obvious in more complex mechanisms, but can be readily determined by the automated mathematical procedure summarized in Appendix A.

Larger mechanisms can have emergent conservation not detectable from static structure that couples species in more complex ways than in the minimal example in equations 1-3. While merging proportional reactions does not create or destroy invariants of kinetic origin, it can be used to reveal them. We use this principle to identify additional emergent kinetic invariants in an automated procedure, combining two fundamental concepts from chemistry and mathematics: the law of mass action (Guldberg & Waage, 1879) and the fundamental theorem of linear algebra (Strang, 1993). We use the law of mass action to identify and merge reactions with proportional rates. We then use the fundamental theorem of linear algebra to identify the resulting number of emergent invariants (Appendix A).

## 3 Reconciling data-driven discovery with a theoretical basis for emergent conservation

Emergent conservation can be discovered by a purely data-driven method (Liu et al., 2024). Liu et al. (2024) applied a data-driven conservation law discovery method to modeled concentration timeseries from the Julia photochemical model (JPM), a simplified mechanism of ozone formation (Sturm & Silva, 2024; Sturm & Wexler, 2022). The version of JPM used in that work conserved carbon and nitrogen atoms, so it was expected that the data-driven method would recover exactly two conserved quantities. Instead it found three. Blokhuis et al. (2025) later connected the extra invariant in the original mechanism to mass action kinetics, as a case of their coproduction law for chemical reaction networks that relates the number of species, reactions, invariants, and null cycles (Blokhuis et al., 2025, 2026).

We add several processes to JPM to rule out plausible stoichiometric explanations for the emergent invariant and find that it is robust in timeseries trajectories. The updated JPM v1.2 mechanism is shown in Figure 2a. JPMv1.2 is a 16-species, 15-reaction system that is highly constrained by design, conserving atoms of all four elements (C, N, H, and O) that make up the species composition. We add two additional reactions in v1.2: Ozone loss to hydroxyl radical $O_3$+OH adds an odd oxygen sink, and as in other reactions in JPM, avoids unphysical exact conservation of the expanded odd oxygen family $O_y$ (Bates & Jacob, 2020). A minor loss pathway for peroxyacetyl nitrate PAN+OH connects all carbon in the system, which avoids a structural carbon subpool stoichiometric invariant akin to a conserved moiety (Famili & Palsson, 2003; Reder, 1988). The full mechanism and simulation details are in Supporting Information S3.

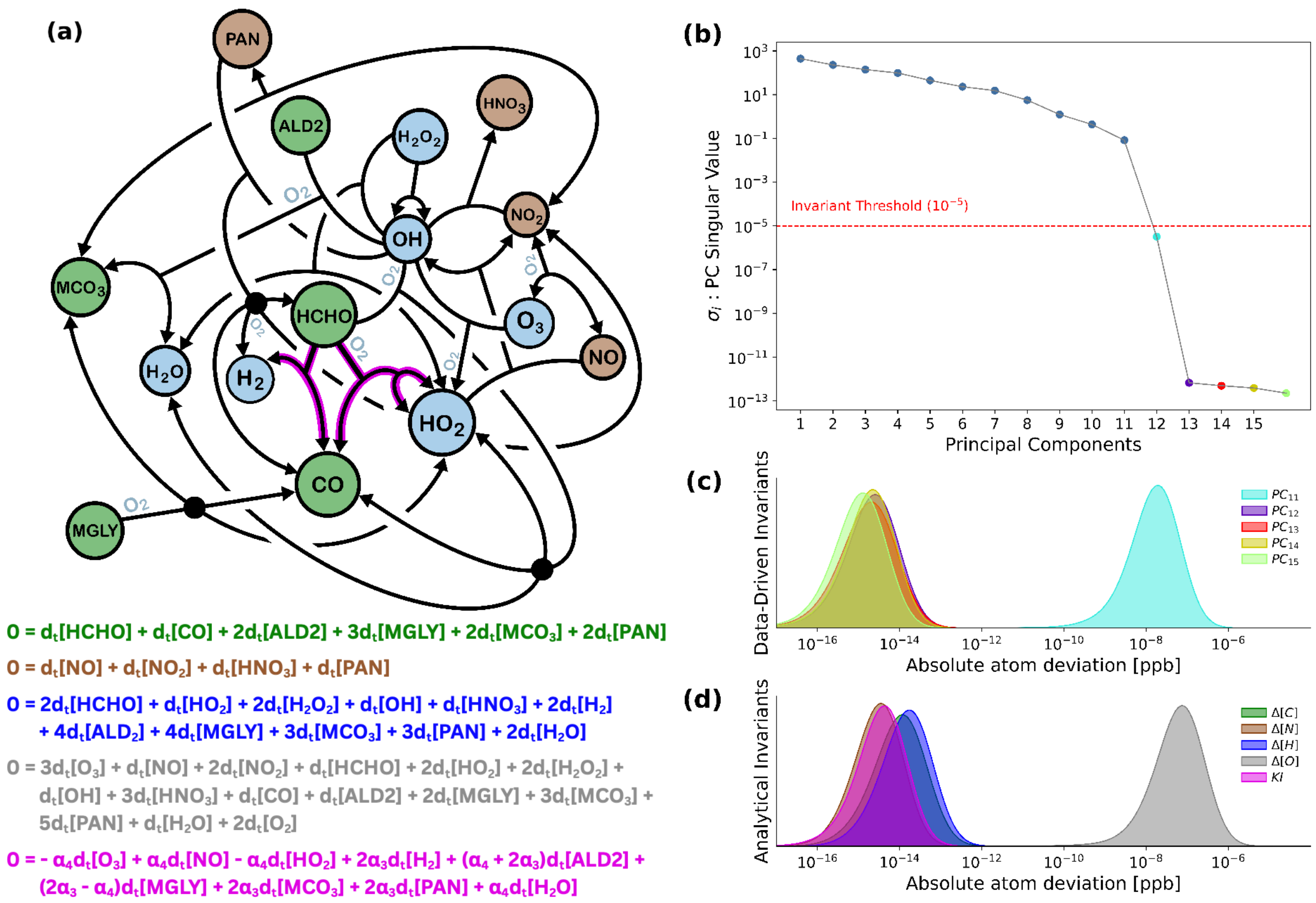

**Figure 2.** a) The Julia photochemical mechanism v1.2 and its invariants, visualized as a species hypergraph inspired by Blokhuis et al. (2025), b) vanishing singular values from principal component analysis on 1,000 simulated concentration timeseries from JPM v1.2 c) the extent of conservation of the corresponding eigenvectors of the vanishing principal components d) the extent of conservation of analytical invariants, for which the data-driven vanishing eigenvectors form a basis.

We apply a data-driven method to concentration timeseries from JPM v1.2 similar to Liu et al. (2024), using principal component analysis (PCA) to enumerate directions of minimum variance as eigenvectors whose singular values vanish (i.e., have extremely small values). PCA on chemical timeseries dates back to the 1960s, though earlier work targeted stoichiometric invariants (Aris & Mah, 1963) or approximate mechanism reduction (Vajda et al., 1985). Five orthogonal directions vanish (Figure 2b), each representing a combination of species concentrations that stays constant regardless of how much the individual concentrations change. Deviations from most invariants remain $<10^{-11}$ ppb (Figure 2c) except for one $<10^{-5}$ ppb, determined by $O_2$ printing precision. Conservation of atoms explains four of the five invariants. The fifth has no stoichiometric explanation.

To explain the emergent fifth conserved quantity we need the theoretical basis from the previous section. Formaldehyde photolysis (magenta reactions in Figure 2a) branches to either $HO_2$, initiating a radical chain, or to $H_2$, inert in this mechanism. Merging the proportional branches reduces the number of reactions in JPM from 15 reactions to 14 effective reactions. That reaction reduction is not itself the invariant: merging reactions can create kinetic invariants, but can also just break null cycles. In this case the proportional kinetics combined with the connectivity of JPMv1.2 yield an emergent conserved quantity, coupling species well beyond the photolysis products. This emergent invariant exists in the mechanism regardless of the merge (as the data-driven analysis found in the timeseries with all 15 reactions represented separately). Merging is useful as it represents kinetic redundancy explicitly in the mechanism, allowing us to symbolically derive the kinetic invariant (Text S3) to obtain a relationship much more interpretable than a dense orthogonal basis vector. The five invariants in JPM are written as conserved combinations of species rates in Figure 2a, where the non-integer weights $\alpha_3$ and $\alpha_4$ in the kinetic invariant correspond to the photolytic branching ratios.

The analytically derived kinetic invariant is conserved to machine precision (Figure 2d), staying <$10^{-11}$ ppb like all stoichiometric invariants besides oxygen conservation at <$10^{-5}$ ppb. The sets of data-driven and analytical invariants form a linear basis for each other: one is a weighted combination of the other. As in the $RO_2$+NO example and section S1, the five invariants reveal that the 16-species system evolves in a latent space with only 11 degrees of freedom. This provides an upper bound for the size of a reduced mechanism with identical behavior: an 11-species mechanism with the five identified invariants could perfectly reconstruct the evolution of all 16 species in JPM at any point in time.

## 4 Survey of emergent conservation in atmospheric chemical mechanisms

We conduct a survey of 35 atmospheric chemical mechanisms with varying complexity, scope, and configurations (see the availability statement and Text S4) to evaluate the prevalence of kinetic invariants. We use the linear algebra procedure (Appendix A) to determine the number of structural invariants for each mechanism, then emergent kinetic invariants assuming mass action kinetics.

We discover kinetic invariants in 15 of the 35 mechanisms, annotated with purple stars in Figure 3. Any mechanism with more species than effective reactions (i.e., the number of reactions after merging), shown in the shaded area under the 1:1 line in Figure 3, is guaranteed to have kinetic invariants. Four of the surveyed mechanisms with kinetic invariants are of this sort: JPM v0.2 and v1.2, GEOS-Chem Hg (GC-Hg, Shah et al., 2021), and the Caltech Isoprene Mechanism CIM v5 (Wennberg et al., 2018).

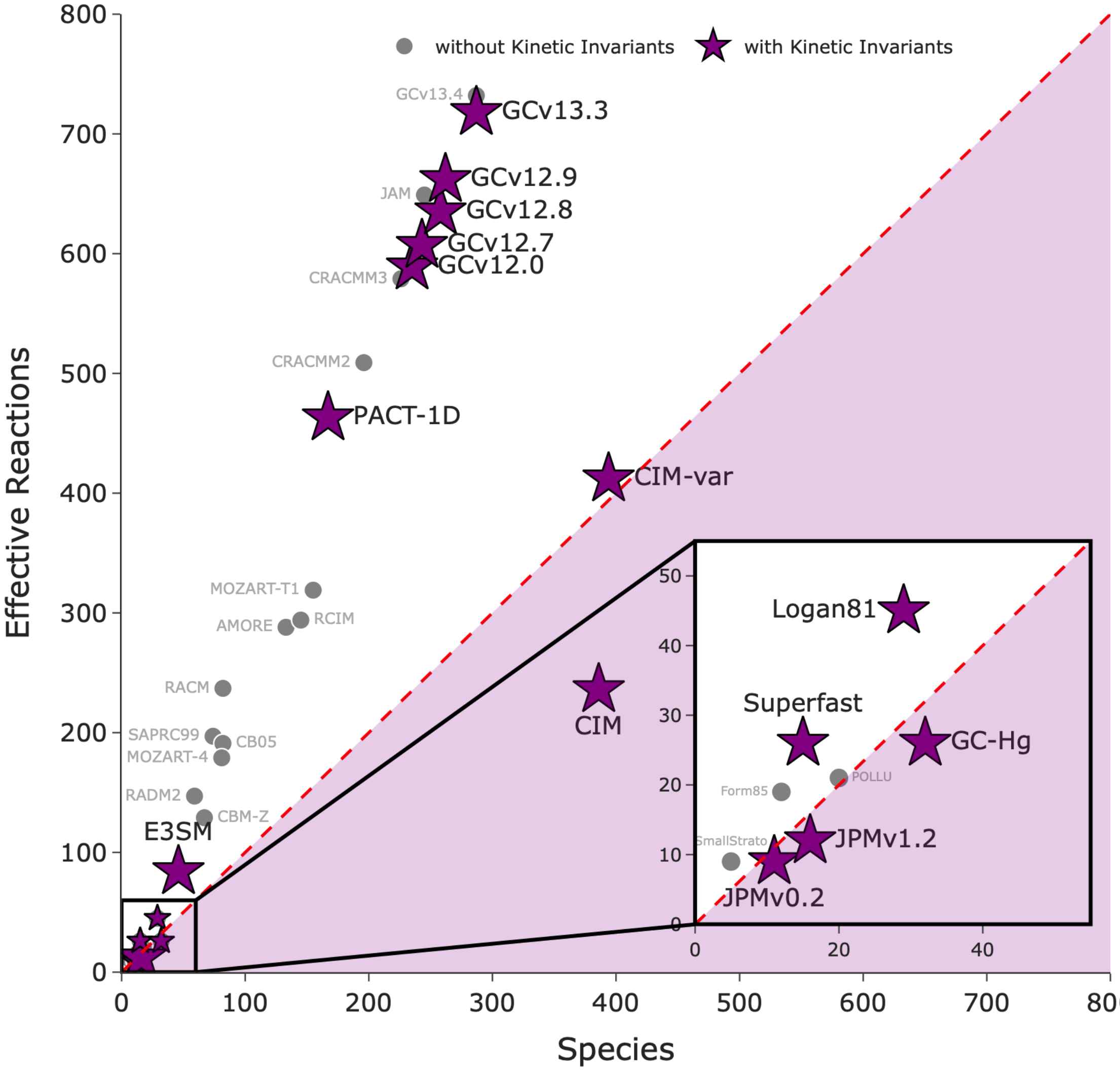

**Figure 3.** A summary of the emergent conservation survey in effective reactions vs species space, plotting 30 of the 35 mechanisms. Mechanisms with emergent conservation are denoted with purple stars. Due to scale constraints, GC14.6.3, CRI, MECCA, MCM, and MCM-PRAM are not shown but are present in Table S2 and in the log-scale Figure S3.

Emergent conservation via kinetic invariants can still exist in mechanisms with more effective reactions than species, of which we find 11 in our survey. The Superfast (Brown-Steiner et al., 2018; Cameron-Smith et al., 2006) and Logan81 (Logan et al., 1981) mechanisms are small enough that it is possible to compute a symbolic representation of all their invariants (Appendix A). For Superfast, the single kinetic invariant is:

$$d_t[SO_2] + d_t[SO_4] + \left(\frac{k_{18} + 0.75k_{19}}{k_{18} + k_{19}}\right) d_t[DMS] = 0 \qquad (6)$$

which corresponds to broken sulfur conservation. The Logan81 mechanism has a single kinetic invariant arising from the branching $N_2O$ reaction with singlet oxygen to form 2 NO or the inert $N_2$. The invariant is:

$$d_t[N_2O] + \left(\frac{k_6 + k_7}{k_6}\right) d_t[N_2] = 0 \qquad (7)$$

A sensitivity test holding $N_2$ concentration constant does not remove the presence of a kinetic invariant in Logan81, though it removes one stoichiometric mass-conserving invariant corresponding to nitrogen conservation.

All versions of the GEOS-Chem mechanism have many proportional reactions leading to a smaller number of effective reactions. This differs from other chemical mechanisms, like CRACMM (Pye et al., 2023; Skipper et al., 2024). While CRACMM3 has many fewer reactions than the mechanism used in GEOS-Chem v12.0, these mechanisms have nearly the same number of effective reactions (579 versus 591) due to the much higher number of proportional reactions in GEOS-Chem. Kinetic invariants can be found in versions of GEOS-Chem through v13.3. GEOS-Chem v12.0, v12.7, and v12.8 each have one. This becomes two kinetic invariants in GCv12.9 through v13.3. With the release of GCv13.4, four sulfate tracers tracking uptake pathways for two hypohalous acids were combined into a single tracer, and the kinetic invariants disappear, remaining absent through GCv14.6.3. The two invariants were each a simple branching relationship between two sulfate tracers, and the mechanism update collapsed all four tracers into one, removing within-halogen distinctions the kinetics never supported. More generally, kinetic invariants can be used as diagnostics to mark kinetically degenerate tracers a mechanism distinguishes on paper but that the kinetics will not separate. While automated detection of redundant tracers could potentially be useful in certain model development settings, kinetic invariants are not limited to this simple case: we find more chemically rich invariants in a similarly sized, though much more explicit isoprene chemical mechanism.

The Caltech Isoprene Mechanism (CIM) has a very high number of proportional reactions, resulting in 236 effective reactions between 386 species (placing CIM under the 1:1 line in Figure 3). This high amount of proportional reaction rates exists because in its default configuration, CIM assigns oxidants $HO_x$, $NO_x$, and $O_3$ to fixed species with constant concentrations. The intended use case of the mechanism in this configuration is not to study the impact of isoprene oxidation on these oxidant concentrations. Rather, CIM is used in this configuration for mechanism intercomparison under identical oxidant conditions. The large number of kinetic invariants in CIM with fixed oxidants bound its intrinsic dimensionality to be smaller than its number of species and reactions might indicate. The 386-species CIM with fixed oxidants evolves with only 236 degrees of freedom. In other words, only 236 species are needed to reconstruct all 386 species exactly, providing an upper bound for the size of a reduced mechanism. Reduced isoprene mechanisms based on CIM such as the 133-species AMORE-Isoprene (Wiser et al., 2025) and the 145-species RCIM (Wennberg et al., 2018) are smaller still, but the gap is narrower than species counts suggest: if used as a baseline for mechanism comparison, this configuration of CIM has an intrinsic dimensionality that places it much closer in size to these reduced mechanisms.

We switch all oxidants in CIM to instead be variable species, leaving only M, $O_2$, and two water tracers fixed. We call this 394-species configuration CIM-var, and find it has 412 effective reactions, placing this mechanism above the 1:1 line of effective reactions to species in Figure 3. Seventeen kinetic invariants still appear in CIM-var. Some of these are simple balances between two products represented as terminal by the mechanism, as in previous examples (e.g., Equations 1-3). Other kinetic invariants in CIM-var reveal more chemically interesting behavior. CIM-var has invariants that include intermediate species as well as terminal species, as seen earlier in the JPM kinetic invariant. Isoprene ozonolysis reactions branch to form two different Criegee intermediates ciMACROO and ciMVKOO, which both react with $H_2O$ to form corresponding products MACR3OH3OOH and MVK3OH3OOH, all four coupled in a kinetic invariant:

$$k_{R527}(d_t[\mathrm{ciMACROO}] + d_t[\mathrm{MACR3OH3OOH}]) \\ -k_{R528}(d_t[\mathrm{ciMVKOO}] + d_t[\mathrm{MVK3OH3OOH}]) = 0$$

This can be recognized as a kinetically invariant partitioning between parallel ozonolysis branches. The invariant carries a memory of the initial ozonolysis branching into the resulting product distribution, with the resulting ratio preserved through the subsequent chemistry of each intermediate.

Branching in the isoprene ozonolysis system is well studied (Nguyen et al., 2016), and the Criegee intermediate example illustrates a chemically intuitive consequence of branching at a single ozonolysis step into parallel pathways. Other kinetic invariants link reactions that are not themselves proportional, but still become coupled through higher-order connectivity. One such kinetic invariant in CIM-var comes from an isoprene nitrate peroxy radical ISOP1OH2OO3CO4N that generates hydroperoxide and carbonyl nitrate terminal products through competing $HO_2$ and NO pathways, each with 3 branching reactions.

These $HO_2$ and NO pathways do not have proportional rates themselves, but their product distribution still obeys an exact kinetic constraint:

$$\begin{aligned}&k_{R753}k_{R755}d_t[\mathrm{ISOP1OH2OOH3CO4N}]\\&+k_{R751}k_{R754}d_t[\mathrm{ISOP1OH2N3CO4N}]\\&-k_{R751}k_{R755}d_t[\mathrm{MVK3CO4N}\ ]=0\end{aligned}$$

linking the hydroperoxide ISOP1OH2OOH3CO4N formed only from the $HO_2$ oxidation pathway with a carbonyl nitrate ISOP1OH2N3CO4N formed only from the NO pathway, with only MVK3CO4N shared between both pathways. In this case, proportional branching within each pathway combined with network connectivity between pathways gives rise to a kinetic constraint that connects them. The $HO_2$ and NO reaction pathways of this isoprene nitrate peroxy radical are not independent. Rather, their product distributions are constrained by the above kinetic invariant that reconnects the two pathways through a shared branching product.

Kinetic invariants can range in complexity from trivial pairwise relationships between inert branching products to more complex groupings of species. A final example from CIM-var links seven isoprene nitrate species through 12 rate constants in a chemically rich kinetic invariant:

$$\begin{aligned}&k_{636}\,d_t[\mathrm{ISOP1N2OH3R4OOH}]+k_{636}\left(\frac{k_{634}+k_{635}}{k_{635}}\right)d_t[\mathrm{ISOP1N2OH34O}]\\&+k_{636}\left(\frac{k_{634}+k_{635}}{k_{635}}\right)\left(\frac{k_{741}+k_{742}}{k_{742}}\right)d_t[\mathrm{ISOP1N2OH3CO4OO}]\\&+k_{636}\left(\frac{k_{634}+k_{635}}{k_{635}}\right)\left(\frac{k_{741}+k_{742}}{k_{742}}\right)\left(\frac{k_{746}+k_{747}}{k_{746}}\right)d_t[\mathrm{ISOP1N2OH3CO4N}]\\&+k_{636}\left(\frac{k_{634}+k_{635}}{k_{635}}\right)\left(\frac{k_{741}+k_{742}}{k_{742}}\right)\left(\frac{k_{748}+k_{749}}{k_{749}}\right)d_t[\mathrm{ISOP1N2OH3CO4OOH}]\\&-k_{633}\,d_t[\mathrm{ISOP1N2R3OH4OOH}]-k_{633}\left(\frac{k_{637}+k_{638}}{k_{638}}\right)d_t[\mathrm{ISOP12O3OH4OOH}]=0\end{aligned}$$

This invariant conserves fluxes through an isoprene nitrate oxidation cascade, where six branching ratios constrain the distribution of intermediate and terminal products as represented by CIM-var. Similar to the Criegee invariant, this kinetic invariant preserves information on earlier branching throughout subsequent generations of oxidation chemistry. This kinetic invariant reveals coupling between species that are separated by multiple reaction steps, unifying the oxidation cascade into one kinetically-defined conserved family.

## 5 Discussion

The law of mass conservation is a well-known invariant property of chemical systems. However, atom conservation arising from reaction stoichiometry is not the only time-invariant behavior that can be detected in chemical mechanisms. We demonstrate that mass action kinetics in a chemical system with branching reactions can lead to additional relationships between species, which belong to a unified class of conserved quantities that we term kinetic invariants.

We use a simplified photochemical mechanism to reconcile two independent approaches for identifying these emergent conserved quantities: a data-driven method that detects all invariants in a simulated timeseries using principal component analysis, and a first-principles approach that leverages the law of mass action and fundamental linear algebra to identify the emergent conserved quantity as a kinetic invariant. The first-principles framework can be generalized to any chemical mechanism with mass action kinetics as in recent work by Blokhuis et al. (2025), except we do not handle 2-reaction null cycles separately. In this work, we use this generalized framework to survey 35 commonly used atmospheric chemical mechanisms spanning orders of magnitude in size.

Emergent conservation from kinetic invariants is not just a property of simplified chemical mechanisms. We discover emergent kinetic invariants in 15 out of 35 surveyed mechanisms, including the Caltech Isoprene Mechanism CIM (Wennberg et al., 2018). CIM represents isoprene oxidation in near-explicit detail and is one of the more complex mechanisms in the survey. CIM also contains the largest number of kinetic invariants even in its variable oxidants configuration. While the simplest kinetic invariants describe pairwise balances between products, more complex examples reveal hidden relationships that preserve branching information over multiple oxidation steps and unify divergent product channels into one conserved, kinetically-defined oxidation family. We additionally find that higher-order structure in chemical mechanisms can link pathways that do not have proportional reaction rates.

The present work is limited to exact kinetic invariants that emerge in numerical chemical mechanisms. The exact invariance requirement could be relaxed to identify approximately conserved kinetic groups of species, which could potentially help define an objective basis for the definitions of existing chemical families such as odd oxygen, or perhaps identify additional tightly coupled chemical families that have not yet been recognized. In considering the implications of approximate invariants alongside mechanism incompleteness, we recognize that invariants which are exact within a given mechanism may themselves be approximate when unresolved minor pathways are added. Much incompleteness in a chemical mechanism comes from omitting processes expected to be unimportant to its use cases (though other decisions like process uncertainty can also play a role), so approximate invariants could characterize the extent to which chemical relationships are preserved as mechanisms are simplified or expanded.

These ideas also motivate experimental applications. Chamber experiments are a natural setting for kinetic invariants, as often small numbers of precursors are studied in highly controlled chemical and environmental conditions to constrain rate constants and other kinetic parameters (Nguyen et al., 2023). Chamber studies already leverage fixed yield ratios of products to establish or rule out branched pathways, and these controlled conditions are also where less obvious kinetic invariants might be most detectable. Challenges in detecting emergent conservation in chamber experiments could arise from measurement noise, wall loss or dilution loss that violate closed system assumptions, which complicate mass conservation diagnostics as well. However, disagreement of measurements with theoretically expected kinetic invariants could be a useful way of identifying the presence of unresolved processes. Kinetic invariants could also be used as an alternative way of constraining kinetic parameters. For example, the organic nitrate experiments in Barber et al. (2024) use a chemical system that contains several branching reactions: resulting invariants could in principle constrain the time evolution of the measured system to a single direction from which branching ratios could be obtained. In both models and measurements, kinetic invariants offer new routes for bounding system complexity and revealing underlying relationships in atmospheric chemistry.

## Appendix A: General method for identifying emergent conservation from kinetics

This section briefly outlines the general linear algebraic procedure to detect emergent kinetic invariants. We begin with the stoichiometric matrix of a chemical mechanism. The stoichiometric matrix is a table of values containing species as rows and reactions as columns, with negative and positive values representing how much each species is consumed or produced in each reaction (e.g. Sturm & Wexler, 2020).

We obtain the number of structural invariants by calculating the dimension of the cokernel, also known as the left null space, of the stoichiometric matrix. Most of these are stoichiometric in origin, but we note that structural invariants with non-integer weights are likely invariants of kinetic origin from branching reactions that have already been merged with fixed ratios (rather than symbolic functions of environmental conditions). Here, our focus is on emergent conservation: the remaining kinetic invariants invariants hidden to the cokernel of the stoichiometric matrix. To account for any remaining kinetic coupling from branching, we merge reactions with duplicate co-reactant sets and therefore proportional rates, using rate constants to proportionally weight each reaction in the merge with pseudo-stoichiometric, environmentally-dependent coefficients corresponding to branching ratios. To find the total number of invariants, we then recompute the dimensionality of the cokernel of the merged matrix (such as the matrix on the right side of Figure 1) – symbolically for small mechanisms and numerically for larger mechanisms (further details in section S4). As the merging operations preserve mass balances and all stoichiometric invariants, any resulting addition to the dimension of the cokernel of the merged matrix is a result of emergent conservation arising from kinetic invariants.

## Acknowledgments

This work was supported by NSF Grant AGS-2228923, NSF Grant AGS-2441535, and NASA Grant 80NSSC23K0523. Obin Sturm acknowledges funding from the Diane Sonosky Montgomery and Jerol Sonosky Graduate Fellowship for Environmental Sustainability Research. We thank Bob Yantosca and Rolf Sander for reviewing and supporting the updates to the Kinetic PreProcessor.

## Availability Statement

As part of this work, we publish a graph library of oxidative atmospheric mechanisms, available at https://doi.org/10.5281/zenodo.19928697 (Sturm et al., 2026). This library contains edge lists of 35 chemical mechanisms in the species-reaction graph format (Feinberg, 2019; Silva et al., 2021). As many mechanisms are written using the Kinetic PreProcessor KPP (Damian et al., 2002; Lin et al., 2023) we formalize the modifications made to KPP to create the species-reaction edge list for the GEOS-Chem mechanism in prior work (Getter et al., 2025; Li et al., 2025). These updates to the KPP codebase are released as part of KPP 3.4.0 with the option to automatically generate the edge list of the species-reaction graph for any chemical mechanism. All other scripts and data used in the analysis are available at 10.5281/zenodo.19928145 (Rodriguez & Sturm, 2026).

## Conflict of Interest Disclosure

The authors declare there are no conflicts of interest for this manuscript.

Supporting Information for

# Emergent conservation in atmospheric chemical mechanisms

**Beatriz Lucia G. Rodriguez[1]†, Patrick Obin Sturm[1]†, Daniel Getter[1], and Sam J. Silva[1,2,3]**

[1]Department of Earth Sciences, University of Southern California, Los Angeles, CA, USA

[2]Department of Environmental Engineering, University of Southern California, Los Angeles, CA, USA

[3]Department of Population and Public Health Sciences, University of Southern California, Los Angeles, CA, USA

Corresponding author: Patrick Obin Sturm (psturm@usc.edu)

†Equally contributing authors

**Contents of this file**

**Introduction**

This contains the supporting information and figures for above manuscript. Table S2 details the results of the 35-mechanism survey and includes source notes.

**Text S1. $RO_2$+NO branching reactions example: deriving a kinetic invariant and a reduced system with identical behavior**

The following section demonstrates how a kinetic invariant emerges in a small chemical system ($RO_2$+NO branching reactions), and how invariants constrain the space in which this chemical system evolves. We start with 5 species of interest interacting in 3 reactions:

(R1) $RO_2 + NO \rightarrow RO + NO_2$

(R2) $RO_2 + NO \rightarrow RONO_2$

(R3) $RO \overset{O_2}{\rightarrow} RO_2$

This system can be illustrated by a 5-by-3 stoichiometric matrix on the left side of Figure S1, where columns represent the reactions and each entry represents stoichiometric coefficients (-x or +x for reactants or products, respectively). Each reaction column is balanced with respect to nitrogen and the general "R" group (e.g. a hydrocarbon chain or an aromatic ring). This means that with the assumption of a closed system, the total number of nitrogen atoms and the number of R groups in the system is constant. Note that the balance of oxygen is not tracked in this example.

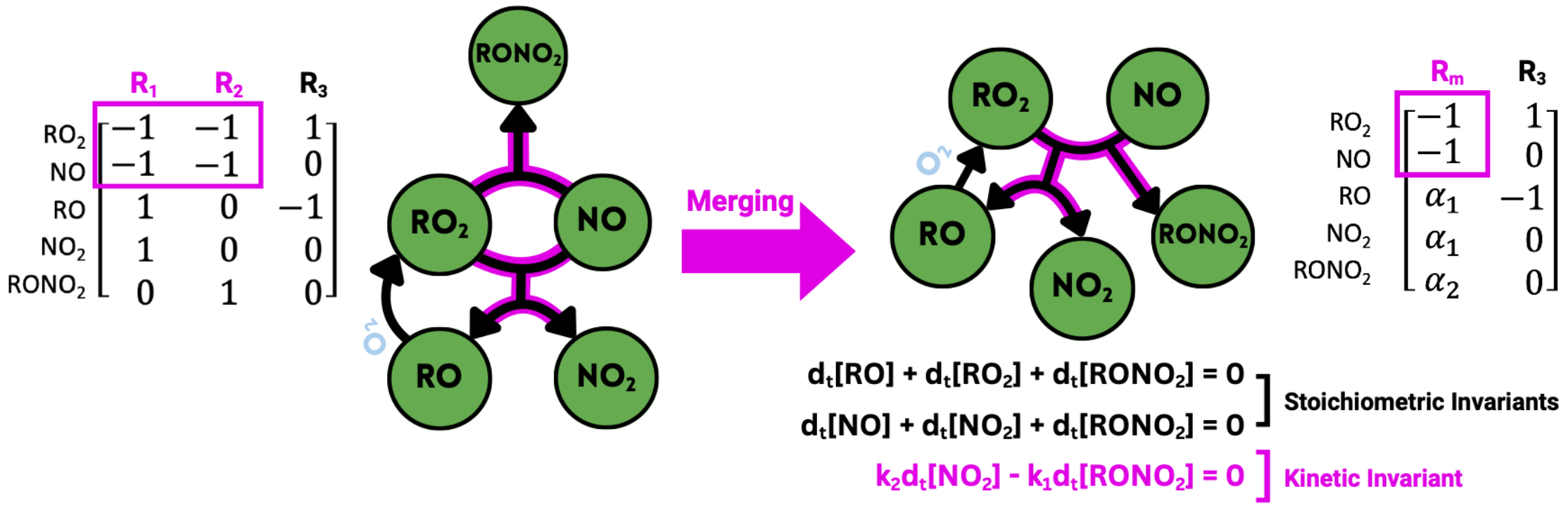

**Figure S1.** Merging branching reactions in the $RO_2$+NO system reveals a kinetic invariant.

By the law of mass action, each reaction proceeds at rates proportional to the product of the concentrations of the co-reactants via rate constant proportionalities $k_1, k_2, k_3$, respectively. The rates of change of each species can then be determined by which reactions it is consumed or produced in. For example, $RO_2$ is consumed in R1 and R2, and produced in R3:

I. $RO_2$ is consumed at the rate $k_1[RO_2][NO]$ in R1

II. $RO_2$ is consumed at the rate $k_2[RO_2][NO]$ in R2

III. $RO_2$ is produced at the rate $k_3[RO]$ in R3

Denoting time rates of change as $d_t$ we then obtain

(E1) $d_t[RO_2] = -k_1[RO_2][NO] - k_2[RO_2][NO] + k_3[RO]$

For the remaining 4 species in the system, we follow the same procedure to get:

(E2) $d_t[NO] = -k_1[RO_2][NO] - k_2[RO_2][NO]$

(E3) $d_t[RO] = k_1[RO_2][NO] - k_3[RO]$

(E4) $d_t[NO_2] = k_1[RO_2][NO]$

(E5) $d_t[RONO_2] = k_2[RO_2][NO]$

The species in the system are chemical combinations of the elements O and N, and the R group. Since the overall concentrations of N and R are conserved within the system by mass balance, their concentration changes can be mathematically represented by these 2 stoichiometric invariants:

Change in concentration of R-containing species: $d_t[RO] + d_t[RO_2] + d_t[RONO_2] = 0$

Change in concentration of N-containing species: $d_t[NO] + d_t[NO_2] + d_t[RONO_2] = 0$

Integrating each with respect to time produces 2 conservation laws $L_1$and $L_2$:

$L_1 = [RO_2] + [RO] + [RONO_2]$, summing the concentrations of R-containing species

$L_2 = [NO_2] + [NO] + [RONO_2]$, summing the concentrations of N-containing species

Notice that in the original system (written below as stoichiometric matrix (M1)), columns 1 and 2 have the same reactant species $RO_2$ and $NO$, i.e. a pair of branching reactions. These reactions can be "merged" within the matrix, with the law of mass action kinetics leading to proportional rates for each of the 2 different product channels. The original stoichiometric coefficients are replaced by $\alpha_1 = \frac{k_1}{k_1+k_2}$ and $\alpha_2 = 1 - \frac{k_1}{k_1+k_2} = \frac{k_2}{k_1+k_2}$ for species produced in reaction 1 and reaction 2 at new fractional kinetic rates $\alpha_1$ and $\alpha_2$ respectively. The newly merged effective reaction (R4) is shown in the "merged" matrix (M2) below (and the right side of Figure S1) and proceeds at a rate of $(k_1 + k_2)\ [RO_2][NO]$. In this simplified system of 5 species and 3 reactions, coupled kinetics thus leads to only 2 effective reactions.

(R4) $RO_2 + NO \rightarrow \alpha_1(RO + NO_2) + (1 - \alpha_1)RONO_2$

$$\text{(M1)}\quad \begin{matrix} RO_2 \\ RO \\ NO \\ NO_2 \\ RONO_2 \end{matrix} \begin{bmatrix} -1 & -1 & 1 \\ 1 & 0 & -1 \\ -1 & -1 & 0 \\ 1 & 0 & 0 \\ 0 & 1 & 0 \end{bmatrix} \rightarrow \quad \text{(M2)}\quad \begin{matrix} RO_2 \\ RO \\ NO \\ NO_2 \\ RONO_2 \end{matrix} \begin{bmatrix} -1 & 1 \\ \alpha_1 & -1 \\ -1 & 0 \\ \alpha_1 & 0 \\ \alpha_2 & 0 \end{bmatrix}$$

The kinetics of the branching reactions in this system lead to an emergent conserved property, a kinetic invariant $L_3$. This arises because products of the two merged reactions, $NO_2$ and $RONO_2$, are linked and kinetically proportional. Taking the rates of change of the products in (E4) and (E5) and rearranging to solve for $[RO_2][NO]$, we obtain

(E6) $[RO_2][NO] = \frac{d_t[NO_2]}{k_1}$

(E7) $[RO_2][NO] = \frac{d_t[RONO_2]}{k_2}$

Equating (E6) and (E7) produces $\frac{d_t[NO_2]}{k_1} = \frac{d_t[RONO_2]}{k_2}$ , or $k_2 * d_t[NO_2] = k_1 * d_t[RONO_2]$, revealing a relation between the two rates of change. Rearranging once more, we identify the emergent kinetic invariant, or the specific combination of rates that is always zero:

$$k_2 d_t[NO_2] - k_1 d_t[RONO_2] = 0$$

and integrating results in the integral form of the kinetic invariant, which shows that the specific combination of species concentrations is always a constant in time:

$$L_3 = k_2[NO_2] - k_1[RONO_2]$$

$L_3$ is constant under constant environmental conditions, as it is a function of rate constants which in turn are dependent on temperature (outside of this example, rate constants can be dependent on other environmental conditions: effective-second-order rate constants for termolecular reactions are also dependent on pressure, and photolytic rate constants are a function of actinic flux). In general, the integral form of kinetic invariants can emerge in situations where environmental conditions are held constant but chemical conditions can still be rapidly evolving, such as a chemistry operator step in a 3D atmospheric model, or in a controlled chamber experiment.

With the new conserved quantity $L_3$, we can now reduce the dimensionality of the original system. In other words, as opposed to tracking the concentration of all species, it is possible to only track that of 2 out of 5 species and use the 3 conservation laws to determine the concentrations of the rest.

We can represent the system of the 3 equations for $L_1$, $L_2$, and $L_3$ in matrix form:

(E8) $$\begin{bmatrix} 1 & 1 & 0 & 0 & 1 \\ 0 & 0 & 1 & 1 & 1 \\ 0 & 0 & 0 & k_2 & -k_1 \end{bmatrix} \begin{bmatrix} RO_2 \\ RO \\ NO \\ NO_2 \\ RONO_2 \end{bmatrix} = \begin{bmatrix} L_1 \\ L_2 \\ L_3 \end{bmatrix}$$

We label the first L.H.S. matrix $M^T$ (where $M$ is the matrix form of the 3 conservation laws). The second L.H.S. matrix we call $C$, the matrix of concentration values for each species.

(E9) $$M^T C = L = \begin{bmatrix} L_1 \\ L_2 \\ L_3 \end{bmatrix}$$

We can look at $M^T$ as a combination of two matrices $M_{free}$ and $M_{dep}$, determined by which species have concentrations that are dependent on or constrained by chemical kinetics. We choose $[RO_2], [RO]$ to be the "free" species (though this is not the only combination of species that can be chosen to be free).

(E10) $$M^T = [M_{free}|M_{dep}]$$

(E11) $$\left[M_{free}\middle|zeros\right] \begin{bmatrix} RO_2 \\ RO \\ NO \\ NO_2 \\ RONO_2 \end{bmatrix} + \left[zeros\middle|M_{dep}\right] \begin{bmatrix} RO_2 \\ RO \\ NO \\ NO_2 \\ RONO_2 \end{bmatrix} = \begin{bmatrix} L_1 \\ L_2 \\ L_3 \end{bmatrix}$$

(E12) $$M_{free} \begin{bmatrix} RO_2 \\ RO \end{bmatrix} + M_{dep} \begin{bmatrix} NO \\ NO_2 \\ RONO_2 \end{bmatrix} = L$$

Using linear algebra, we isolate and solve for $\begin{bmatrix} NO \\ NO_2 \\ RONO_2 \end{bmatrix}$ from (E12) by subtracting $M_{free} \begin{bmatrix} RO_2 \\ RO \end{bmatrix}$ from $L$ and multiplying the difference by the inverse of $M_{dep}$:

(E13) $$\begin{bmatrix} NO \\ NO_2 \\ RONO_2 \end{bmatrix} = M_{dep}^{-1}(L - M_{free} \begin{bmatrix} RO_2 \\ RO \end{bmatrix})$$

We rearrange $L_1$, $L_2$, and $L_3$ :

(E14) $$[RONO_2] = L_1 - [RO_2] - [RO]$$

(E15) $$[NO] = L_2 - [NO_2] - [RONO_2]$$

(E16) $$[NO_2] = \frac{L_3 + k_1[RONO_2]}{k_2}$$

We can rewrite the rearranged conservation law equations to be in terms of only the free species, chosen to be $RO_2$ and $RO$. We substitute $[RONO_2]$ in (E16) with the R.H.S. of (E14) to get:

(E17) $[NO_2] = \frac{L_3+k_1(L_1-[RO_2]-[RO])}{k_2}$

Substituting (E17)'s R.H.S. into $[NO_2]$, and (E14)'s R.H.S. into $[RONO_2]$ in (E15) produces:

(E18) $[NO] = L_2 - \frac{L_3+k_1(L_1-[RO_2]-[RO])}{k_2} - (L_1 - [RO_2] - [RO])$

Now we perform algebraic substitutions and transformations to reduce the system of equations.

From (E1) we have:

$$d_t[RO_2] = -k_1[RO_2][NO] - k_2[RO_2][NO] + k_3[RO]$$

$$d_t[RO_2] = -(k_1 + k_2)[RO_2][NO] + k_3[RO]$$

We substitute the R.H.S. of (E18) into [NO]:

$$d_t[RO_2] = -(k_1 + k_2)[RO_2](L_2 - \frac{L_3 + k_1(L_1 - [RO_2] - [RO])}{k_2} - (L_1 - [RO_2] - [RO])) + k_3[RO]$$

$$d_t[RO_2] = -(k_1 + k_2)[RO_2](L_2 - \frac{L_3}{k_2} - (\frac{k_1 + k_2}{k_2})L_1) - (k_1 + k_2)[RO_2](\frac{k_1 + k_2}{k_2})([RO_2] + [RO]) + k_3[RO]$$

$$d_t[RO_2] = -\frac{(k_1 + k_2)^2}{k_2}[RO_2]^2 - \frac{(k_1 + k_2)^2}{k_2}[RO_2][RO] - (\frac{k_1 + k_2}{k_2})(k_2L_2 - L_3 - (k_1 + k_2)L_1)[RO_2] + k_3[RO]$$

From (E3) we have:

$$d_t[RO] = k_1[RO_2][NO] - k_3[RO]$$

Similarly, we substitute [NO]:

$$d_t[RO] = k_1[RO_2](L_2 - \frac{L_3 + k_1(L_1 - [RO_2] - [RO])}{k_2} - (L_1 - [RO_2] - [RO])) - k_3[RO]$$

$$d_t[RO] = k_1[RO_2](L_2 - \frac{L_3}{k_2} - \frac{k_1(L_1)}{k_2} + \frac{k_1[RO_2]}{k_2} + \frac{k_1[RO]}{k_2} - (L_1 - [RO_2] - [RO])) - k_3[RO]$$

$$d_t[RO] = k_1[RO_2](L_2 - \frac{L_3}{k_2} - L_1\frac{k_1 + k_2}{k_2} + [RO_2]\frac{k_1 + k_2}{k_2} + [RO]\frac{k_1 + k_2}{k_2}) - k_3[RO]$$

$$d_t[RO] = \frac{k_1(k_1+k_2)}{k_2}[RO_2]^2 + \frac{k_1(k_1+k_2)}{k_2}[RO_2][RO] + (\frac{k_1}{k_2})(k_2L_2 - L_3 \\ - (k_1+k_2)L_1)[RO_2] - k_3[RO]$$

We end up with the following two differential equations, reducing the system of originally 5 dimensions to only 2.

$$d_t[RO_2] = -\frac{(k_1+k_2)^2}{k_2}[RO_2]^2 - \frac{(k_1+k_2)^2}{k_2}[RO_2][RO] - (\frac{k_1+k_2}{k_2})(k_2L_2 - L_3 \\ - (k_1+k_2)L_1)[RO_2] + k_3[RO]$$

$$d_t[RO] = \frac{k_1(k_1+k_2)}{k_2}[RO_2]^2 + \frac{k_1(k_1+k_2)}{k_2}[RO_2][RO] + (\frac{k_1}{k_2})(k_2L_2 - L_3 \\ - (k_1+k_2)L_1)[RO_2] - k_3[RO]$$

This 2-dimensional ODE system evolves in exactly the same way as the 5-dimensional ODE system, under fixed environmental conditions. In other words, the system of 5 species evolves on a 2D manifold within a 5D space. As derived above, there are 3 constraints on the system: the first 2 are a consequence of mass conservation of N atoms and the R group respectively, and the third arises from emergent conservation. Given the constants $L_1, L_2$ and $L_3$, the concentrations of the 3 other species can be reconstructed from $[RO_2]$ and $[RO]$ at any point. We visualize the intersection of these constraints in rate form, parallel to Figure 1 and projected onto species constrained just by conservation of N.

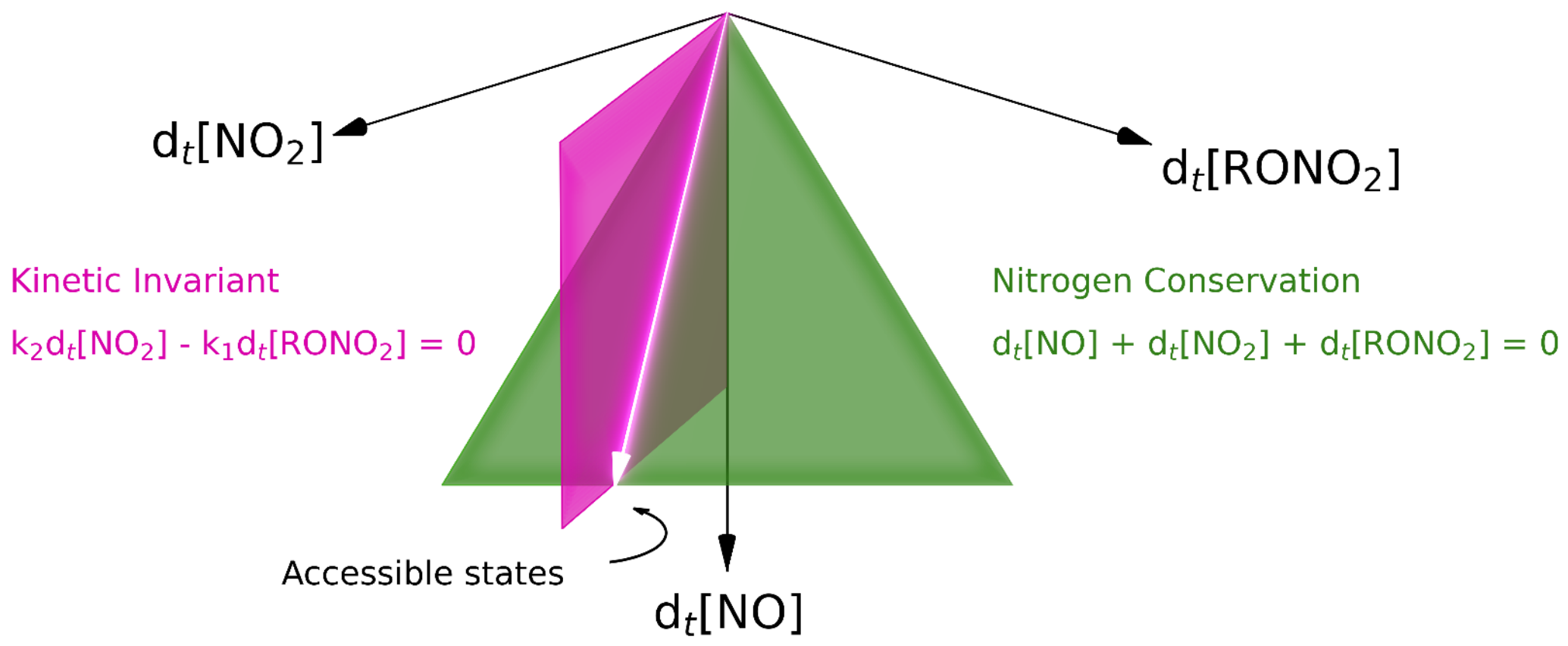

**Figure S2.** Stoichiometric and kinetic invariants (shown here in rate form) limit a system's accessible states to their intersection.

## Text S2. Demonstration of a broken null cycle in the Chapman mechanism

Branching reactions do not always produce an emergent invariant like in the $RO_2 + NO$ example. In this section we demonstrate how this can result in a broken null cycle instead (Blokhuis et al., 2025). The Chapman mechanism (Chapman, 1932), with 4 species of interest and 5 reactions, has one instance of coproduction: reactions (R3) and (R4). See the below set of reactions, and the matrices in Figure S1.

(R1) $\quad O_2 \xrightarrow{hv} 2O(^3P)$

(R2) $\quad O(^3P) + O_2 \xrightarrow{M} O_3$

(R3) $\quad O_3 \xrightarrow{hv} O(^3P) + O_2$

(R4) $\quad O_3 \xrightarrow{hv} O(^1D) + O_2$

(R5) $\quad O(^1D) \xrightarrow{M} O(^3P)$

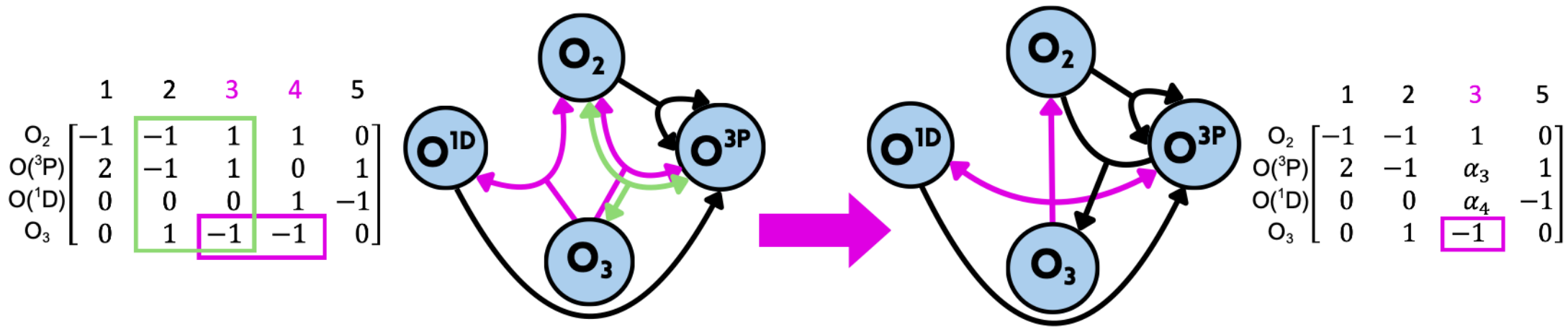

**Figure S3.** A broken null cycle from coproduction in the Chapman mechanism

(M1) is the stoichiometric matrix of the Chapman mechanism, and (M2) is the result of merging reactions (R3) and (R4). Following the merging procedure, we see that $O(^3P)$ and $O(^1D)$ are produced at the fractional kinetic rates $\alpha_3$ and $\alpha_4$ respectively.

$$\text{(M1)} \quad \begin{matrix} O_2 \\ O(^3P) \\ O(^1D) \\ O_3 \end{matrix} \begin{bmatrix} -1 & -1 & 1 & 1 & 0 \\ 2 & -1 & 1 & 0 & 1 \\ 0 & 0 & 0 & 1 & -1 \\ 0 & 1 & -1 & -1 & 0 \end{bmatrix} \rightarrow \quad \text{(M2)} \quad \begin{matrix} O_2 \\ O(^3P) \\ O(^1D) \\ O_3 \end{matrix} \begin{bmatrix} -1 & -1 & 1 & 0 \\ 2 & -1 & \alpha_3 & 1 \\ 0 & 0 & \alpha_4 & -1 \\ 0 & 1 & -1 & 0 \end{bmatrix}$$

The newly merged reaction can be written as

(R6) $\quad O_3 \xrightarrow{hv} \alpha_3 O(^3P) + \alpha_4 O(^1D) + O_2$

Notice that (R2) and (R3) together make up a null cycle (the net change of the respective participating species is zero when both reactions occur simultaneously). However, (R3) also participates in coproduction. One unit of $O(^3P)$ reacts with $O_2$ in (R2) to complete the production of ozone for consumption in (R3), but because of coproduction, only $\alpha_3$-amount of $O(^3P)$ is produced and available to react at a time, hence the disruption of the null cycle. From a linear algebra perspective, an emergent invariant is usually identified by a reduction in rank from (M1) to (M2). But null cycles create a set of linearly dependent vectors that prevent the matrix from being full-rank in the first place. Breaking the null cycle cancels out the rank-reduction effect of merging and thus doesn’t lead to the appearance of an emergent invariant. The merged matrix (M2) (also depicted on the right side of Figure S1) helps us visualize the loss of the cycle. The ranks of both (M1) and (M2) can be verified to be the same – both equal to 3.

**Text S3. Derivation of the kinetic invariant in the Julia photochemical mechanism**

This section demonstrates how a kinetic invariant emerges in the Julia Photochemical Mechanism JPM v1.2. Because of its simplicity and highly constrained nature, JPM has been used to prototype methods for exploration of chemical mechanism behavior and data-driven emulation (Guo et al., 2024; Kircher & Votsmeier, 2025; Liu et al., 2024; Yang et al., 2024).

From JPM v1.1, we add two reactions to the mechanism to avoid unphysical stoichiometric laws as explained in Section 3. For input to the data-driven method, we run 1000 1-hour simulations with the same timesteps and randomly initialized ranges of concentrations as the JPM v1.1 experiments detailed in Sturm and Silva (2024). We do not include other processes (e.g. emissions) to isolate gas-phase chemical processes specifically. For these experiments we vary temperature between 283 and 303 K, keeping all other environmental conditions the same as before. As in prior experiments, $O_2$ has been added to track oxygen conservation, though it does not participate as an explicit factor in the otherwise mass-action rate laws. The updated mechanism with 16 species interacting in 15 reactions is shown in Table S1.

**Table S1.** The Julia photochemical model version 1.2

| Reaction Number | Reaction | Rate constant reference |
|---|---|---|
| R1 | $NO_2 + O_2 + hv \rightarrow NO + O_3$ | See note 1 |
| R2 | $O_3 + NO \rightarrow NO_2 + O_2$ | JPL 19-5 (Burkholder et al., 2020) page 84 Table 1C |
| R3 | $HCHO + 2O_2 + hv \rightarrow 2\ HO_2{\cdot} + CO$ | See note 1 |
| R4 | $HCHO + hv \rightarrow H_2 + CO$ | See note 1 |
| R5 | $HCHO + HO{\cdot} + O_2 \rightarrow HO_2{\cdot} + CO + H_2O$ | JPL 19-5 (Burkholder et al., 2020) page 106 Table 1D |
| R6 | $HO_2{\cdot} + NO \rightarrow OH{\cdot} + NO_2$ | JPL 19-5 (Burkholder et al., 2020) page 83 Table 1C |
| R7 | $OH{\cdot} + NO_2 \rightarrow HNO_3$ | See note 2, JPL 19-5 (Burkholder et al., 2020) page 434 Table 2-1 |
| R8 | $HO_2H + hv \rightarrow 2\ OH{\cdot}$ | See note 1 |
| R9 | $HO_2H + OH{\cdot} \rightarrow H_2O + HO_2{\cdot}$ | T-independent recommendation, JPL 19-5 (Burkholder et al., 2020) page 73 |
| R10 | $ALD2 + OH + O_2 \rightarrow MCO_3 + H_2O$ | JPL 19-5(Burkholder et al., 2020) page 107 Table 1D |
| R11 | $MGLY + 2O_2 \rightarrow MCO_3 + CO + HO_2{\cdot}$ | See note 1 |
| R12 | $MCO_3 + NO_2 \rightarrow PAN$ | See note 2, JPL 19-5 (Burkholder et al., 2020) page 435 Table 2-1 |
| R13 | $PAN \rightarrow MCO_3 + NO_2$ | Equil. Constant / k12, JPL 19-5 (Burkholder et al., 2020) page 521 Table 3-1 |
| R14 | $PAN + OH \rightarrow HCHO + CO + NO_2 + H_2 + O_2$ | Taken from GEOS-Chem v14.7.0 and inert products to balance, also see note 3 |
| R15 | $O_3 + OH \rightarrow HO_2 + O_2$ | JPL 19-5 (Burkholder et al., 2020) page 67 table 1B reaction B6 |

Note 1: Photolysis rate taken from the GEOS-CF full chemical state sampled using the KPP Standalone Interface at a grid cell over Los Angeles at local noon (Keller et al., 2021; Li et al., 2025)
Note 2: Termolecular reaction with effective second-order rate constant calculated from low pressure and high pressure limits and total gas concentration (Burkholder et al., 2020)
Note 3: This minor loss pathway rate constant for PAN was taken from GEOS-Chem v14.7.0, 3.00d-14 is less than the JPL 19-5 recommendation of <4.0d-14 in reaction D46

We derive the kinetic invariant with similar mathematical procedures as that of the sample $RO_2$ mechanism. The JPM mechanism is represented as a 16-by-15 stoichiometric matrix (M1) below, with rows corresponding to species and columns corresponding to reactions.

$(M_1)$

| $spc$ | R1 | R2 | R3 | R4 | R5 | R6 | R7 | R8 | R9 | R10 | R11 | R12 | R13 | R14 | R15 |
|---|---|---|---|---|---|---|---|---|---|---|---|---|---|---|---|
| $O_3$ | 1 | −1 | 0 | 0 | 0 | 0 | 0 | 0 | 0 | 0 | 0 | 0 | 0 | 0 | −1 |
| $NO$ | 1 | −1 | 0 | 0 | 0 | −1 | 0 | 0 | 0 | 0 | 0 | 0 | 0 | 0 | 0 |
| $NO_2$ | −1 | 1 | 0 | 0 | 0 | 1 | −1 | 0 | 0 | 0 | 0 | −1 | 1 | 1 | 0 |
| $HCHO$ | 0 | 0 | −1 | −1 | −1 | 0 | 0 | 0 | 0 | 0 | 0 | 0 | 0 | 1 | 0 |
| $HO_2$ | 0 | 0 | 2 | 0 | 1 | −1 | 0 | 0 | 1 | 0 | 1 | 0 | 0 | 0 | 1 |
| $H_2O_2$ | 0 | 0 | 0 | 0 | 0 | 0 | 0 | −1 | −1 | 0 | 0 | 0 | 0 | 0 | 0 |
| $OH$ | 0 | 0 | 0 | 0 | −1 | 1 | −1 | 2 | −1 | −1 | 0 | 0 | 0 | −1 | −1 |
| $HNO_3$ | 0 | 0 | 0 | 0 | 0 | 0 | 1 | 0 | 0 | 0 | 0 | 0 | 0 | 0 | 0 |
| $CO$ | 0 | 0 | 1 | 1 | 1 | 0 | 0 | 0 | 0 | 0 | 1 | 0 | 0 | 1 | 0 |
| $H_2$ | 0 | 0 | 0 | 1 | 0 | 0 | 0 | 0 | 0 | 0 | 0 | 0 | 0 | 1 | 0 |
| $ALD2$ | 0 | 0 | 0 | 0 | 0 | 0 | 0 | 0 | 0 | −1 | 0 | 0 | 0 | 0 | 0 |
| $MGLY$ | 0 | 0 | 0 | 0 | 0 | 0 | 0 | 0 | 0 | 0 | −1 | 0 | 0 | 0 | 0 |
| $MCO_3$ | 0 | 0 | 0 | 0 | 0 | 0 | 0 | 0 | 0 | 1 | 1 | −1 | 1 | 0 | 0 |
| $PAN$ | 0 | 0 | 0 | 0 | 0 | 0 | 0 | 0 | 0 | 0 | 0 | 1 | −1 | −1 | 0 |
| $H_2O$ | 0 | 0 | 0 | 0 | 1 | 0 | 0 | 0 | 1 | 1 | 0 | 0 | 0 | 0 | 0 |
| $O_2$ | −1 | 1 | −2 | 0 | −1 | 0 | 0 | 0 | 0 | −1 | −2 | 0 | 0 | 1 | 1 |

By construction, this mechanism conserves the atoms of 4 elements: $C$, $H$, $N$, and $O$. Thus we have 4 initial conservation laws.

$$L_C = [HCHO] + [CO] + 2[ALD2] + 3[MGLY] + 2[MCO_3] + 2[PAN]$$

$$L_N = [NO] + [NO_2] + [HNO_3] + [PAN]$$

$$L_H = 2[HCHO] + [HO_2] + 2[H_2O_2] + [OH] + [HNO_3] + 2[H_2] + 4[ALD2] + 4[MGLY] + 3[MCO_3] + 3[PAN] + 2[H_2O]$$

$$L_O = 3[O_3] + [NO] + 2[NO_2] + [HCHO] + 2[HO_2] + 2[H_2O_2] + [OH] + 3[HNO_3] + [CO] + [ALD2] + 2[MGLY] + 3[MCO_3] + 5[PAN] + [H_2O] + 2[O_2]$$

$L_C$, $L_N$, $L_H$, $L_O$ correspond to carbon, nitrogen, hydrogen, and oxygen conservation respectively. These stoichiometric invariants (i.e. structural invariants from atom conservation that exist regardless of kinetics) can be represented as the columns of the species-atom composition matrix $M_L$.

$$
(M_L) \quad
\begin{array}{r}
\\
O_3 \\ NO \\ NO_2 \\ HCHO \\ HO_2 \\ H_2O_2 \\ OH \\ HNO_3 \\ CO \\ H_2 \\ ALD2 \\ MGLY \\ MCO_3 \\ PAN \\ H_2O \\ O_2
\end{array}
\begin{array}{c}
\begin{array}{cccc} \mathrm{C} & \mathrm{N} & \mathrm{H} & \mathrm{O} \end{array} \\
\begin{bmatrix}
0 & 0 & 0 & 3 \\
0 & 1 & 0 & 1 \\
0 & 1 & 0 & 2 \\
1 & 0 & 2 & 1 \\
0 & 0 & 1 & 2 \\
0 & 0 & 2 & 2 \\
0 & 0 & 1 & 1 \\
0 & 1 & 1 & 3 \\
1 & 0 & 0 & 1 \\
0 & 0 & 2 & 0 \\
2 & 0 & 4 & 1 \\
3 & 0 & 4 & 2 \\
2 & 0 & 3 & 3 \\
2 & 1 & 3 & 5 \\
0 & 0 & 2 & 1 \\
0 & 0 & 0 & 2
\end{bmatrix}
\end{array}
$$

The cokernel or left null space of $M_1$, or equivalently the null space of $M_1^T$, is composed of 4 vectors that are the stoichiometric invariants of the mechanism (Blokhuis et al., 2025; Liu et al., 2024).

In this mechanism, we have one instance of proportional rates (neglecting $[O_2]$ in the rate equation): (R3) and (R4) in $M_1$, which have rate constants $k_3$, $k_4$ respectively. The linearly proportional reactions (R3) and (R4), which proceed at reaction rates $r_3$ and $r_4$ respectively, can be represented by a "merge" operation to reaction (Rm) which proceeds at a rate $r_m = r_3 + r_4 = (k_3 + k_4)\,[HCHO]$. We can construct the kinetic-stoichiometric matrix $M_2$ where the original stoichiometric coefficients are replaced by $\alpha_3 = \frac{k_3}{k_3+k_4}$ and $\alpha_4 = 1 - \frac{k_3}{k_3+k_4} = \frac{k_4}{k_3+k_4}$ for species produced in (R3) and (R4) at new branching ratios $\alpha_3$ and $\alpha_4$ respectively.

(Rm) $\quad HCHO \rightarrow \alpha_3(2HO_2 + CO) + (1 - \alpha_3)(H_2 + CO)$

$(M_2)$

| $spc$ | R1 | R2 | Rm | R5 | R6 | R7 | R8 | R9 | R10 | R11 | R12 | R13 | R14 | R15 |
|---|---|---|---|---|---|---|---|---|---|---|---|---|---|---|
| $O_3$ | $1$ | $-1$ | $0$ | $0$ | $0$ | $0$ | $0$ | $0$ | $0$ | $0$ | $0$ | $0$ | $0$ | $-1$ |
| $NO$ | $1$ | $-1$ | $0$ | $0$ | $-1$ | $0$ | $0$ | $0$ | $0$ | $0$ | $0$ | $0$ | $0$ | $0$ |
| $NO_2$ | $-1$ | $1$ | $0$ | $0$ | $1$ | $-1$ | $0$ | $0$ | $0$ | $0$ | $-1$ | $1$ | $1$ | $0$ |
| $HCHO$ | $0$ | $0$ | $-1$ | $-1$ | $0$ | $0$ | $0$ | $0$ | $0$ | $0$ | $0$ | $0$ | $1$ | $0$ |
| $HO_2$ | $0$ | $0$ | $2\alpha_3$ | $1$ | $-1$ | $0$ | $0$ | $1$ | $0$ | $1$ | $0$ | $0$ | $0$ | $1$ |
| $H_2O_2$ | $0$ | $0$ | $0$ | $0$ | $0$ | $0$ | $-1$ | $-1$ | $0$ | $0$ | $0$ | $0$ | $0$ | $0$ |
| $OH$ | $0$ | $0$ | $0$ | $-1$ | $1$ | $-1$ | $2$ | $-1$ | $-1$ | $0$ | $0$ | $0$ | $-1$ | $-1$ |
| $HNO_3$ | $0$ | $0$ | $0$ | $0$ | $0$ | $1$ | $0$ | $0$ | $0$ | $0$ | $0$ | $0$ | $0$ | $0$ |
| $CO$ | $0$ | $0$ | $\alpha_3 + \alpha_4$ | $1$ | $0$ | $0$ | $0$ | $0$ | $0$ | $1$ | $0$ | $0$ | $1$ | $0$ |
| $H_2$ | $0$ | $0$ | $\alpha_4$ | $0$ | $0$ | $0$ | $0$ | $0$ | $0$ | $0$ | $0$ | $0$ | $1$ | $0$ |
| $ALD2$ | $0$ | $0$ | $0$ | $0$ | $0$ | $0$ | $0$ | $0$ | $-1$ | $0$ | $0$ | $0$ | $0$ | $0$ |
| $MGLY$ | $0$ | $0$ | $0$ | $0$ | $0$ | $0$ | $0$ | $0$ | $0$ | $-1$ | $0$ | $0$ | $0$ | $0$ |
| $MCO_3$ | $0$ | $0$ | $0$ | $0$ | $0$ | $0$ | $0$ | $0$ | $1$ | $1$ | $-1$ | $1$ | $0$ | $0$ |
| $PAN$ | $0$ | $0$ | $0$ | $0$ | $0$ | $0$ | $0$ | $0$ | $0$ | $0$ | $1$ | $-1$ | $-1$ | $0$ |
| $H_2O$ | $0$ | $0$ | $0$ | $1$ | $0$ | $0$ | $0$ | $1$ | $1$ | $0$ | $0$ | $0$ | $0$ | $0$ |
| $O_2$ | $-1$ | $1$ | $-2\alpha_3$ | $-1$ | $0$ | $0$ | $0$ | $0$ | $-1$ | $-2$ | $0$ | $0$ | $1$ | $1$ |

Since $CO$ appears as a product in both (R3) and (R4), it is produced in (Rm) with a stoichiometric weight of $(\alpha_3 + \alpha_4)$, which equals 1.

As in the example of the $RO_2 + NO$ branching reactions, the kinetic proportionality of the products of the branching reactions lead to an emergent kinetic invariant $L_{kinetic}$. Note that all stoichiometric and kinetic invariants exist in the mechanism before *and* after merging. The merging operation simply allows us to visualize the conservation laws in a more mathematically interpretable way. The JPM mechanism is larger than the $RO_2$ mechanism so finding $L_{kinetic}$ requires a longer algebraic approach. The overall change in concentration of each species can be written as a sum of the rates of the reactions they participate in. For example, $O_3$ is produced in (R1) and consumed in (R2) and (R15), so the instantaneous time rate of change of $O_3$ concentration can be written as $d_t[O_3] = r_1 - r_2 - r_{15}$, where $r_1$, $r_2$, and $r_{15}$ denote the reaction rates of (R1), (R2), and (R15) respectively.

$$d_t[O_3] = r_1 - r_2 - r_{15}$$

$$d_t[NO] = r_1 - r_2 - r_6$$

$$d_t[HO_2] = 2r_3 + r_5 - r_6 + r_9 + r_{11} + r_{15}$$

$$d_t[H_2] = r_4 + r_{14}$$

$$d_t[ALD2] = -r_{10}$$

$$d_t[MGLY] = -r_{11}$$

$$d_t[H_2O] = r_5 + r_9 + r_{10}$$

$$d_t[PAN] = r_{12} - r_{13} - r_{14}$$

$d_t[MCO_3] = r_{10} + r_{11} - r_{12} + r_{13}$

After the merging of (R3) and (R4) into (Rm), we can represent the rates of $R3$ and $R4$ as ($\alpha_3 r_m$) and ($\alpha_4 r_m$) respectively, affecting only $d_t[HO_2]$ and $d_t[H_2]$.

$d_t[HO_2] = 2\alpha_3 r_m + r_5 - r_6 + r_9 + r_{11} + r_{15}$

$d_t[H_2] = \alpha_4 r_m + r_{14}$

Now we attempt to manually find a vector that fulfills two requirements: 1) it lies in the null space of $M_2$ and 2) it is not already part of the null space of $M_1$. In other words, we want to find a combination of the 9 equations corresponding to species rates that result in net zero change overall. We also aim to make each element of the vector a multiple of $\alpha_3$ or $\alpha_4$ to satisfy requirement (2). We arrive at the following solution:

$$- \alpha_4 * d_t[O_3] = -\alpha_4 r_1 + \alpha_4 r_2 + \alpha_4 r_{15}$$

$$\alpha_4 * d_t[NO] = \alpha_4 r_1 - \alpha_4 r_2 - \alpha_4 r_6$$

$$-\alpha_4 * d_t[HO_2] = -2\alpha_3 \alpha_4 r_m - \alpha_4 r_5 + \alpha_4 r_6 - \alpha_4 r_9 - \alpha_4 r_{11} - \alpha_4 r_{15}$$

$$2\alpha_3 * d_t[H_2] = 2\alpha_3 \alpha_4 r_m + 2\alpha_3 r_{14}$$

$$(\alpha_4 + 2\alpha_3) * d_t[ALD2] = -(\alpha_4 + 2\alpha_3) r_{10}$$

$$(2\alpha_3 - \alpha_4) * d_t[MGLY] = -(2\alpha_3 - \alpha_4) r_{11}$$

$$2\alpha_3 * d_t[MCO_3] = 2\alpha_3 r_{10} + 2\alpha_3 r_{11} - 2\alpha_3 r_{12} + 2\alpha_3 r_{13}$$

$$2\alpha_3 * d_t[PAN] = 2\alpha_3 r_{12} - 2\alpha_3 r_{13} - 2\alpha_3 r_{14}$$

$$\alpha_4 * d_t[H_2O] = \alpha_4 r_5 + \alpha_4 r_9 + \alpha_4 r_{10}$$

This results in the vector $v_1$

$$v_1 = \begin{bmatrix} -\alpha_4 \\ \alpha_4 \\ 0 \\ 0 \\ -\alpha_4 \\ 0 \\ 0 \\ 0 \\ 0 \\ 2\alpha_3 \\ \alpha_4 + 2\alpha_3 \\ 2\alpha_3 - \alpha_4 \\ 2\alpha_3 \\ 2\alpha_3 \\ \alpha_4 \\ 0 \end{bmatrix}$$

corresponding to the kinetic invariant

$$0 = -\alpha_4 d_t[O_3] + \alpha_4 d_t[NO] - \alpha_4 d_t[HO_2] + 2\alpha_3 d_t[H_2] + (\alpha_4 + 2\alpha_3) d_t[ALD2] + (2\alpha_3 - \alpha_4) d_t[MGLY] + 2\alpha_3 d_t[MCO_3] + 2\alpha_3 d_t[PAN] + \alpha_4 d_t[H_2O]$$

which when integrated is equal to a constant in time:

$$L_{kinetic} = -\alpha_4[O_3] + \alpha_4[NO] - \alpha_4[HO_2] + 2\alpha_3[H_2] + (\alpha_4 + 2\alpha_3)[ALD2] + (2\alpha_3 - \alpha_4)[MGLY] + 2\alpha_3[MCO_3] + 2\alpha_3[PAN] + \alpha_4[H_2O]$$

We can verify that adding up all the R.H.S. of the equations with coefficients equals 0. Additionally, performing the matrix operation $v_1^T M_2$ results in the null (zero) vector. Finally, we recognize that $v_1$is linearly independent of the 4 stoichiometric conservation laws $[M_L]$

It is worth noting that, if obtained via other methods, the set of 5 vectors corresponding to the conservation laws might not be the same as our findings. We wanted to see if our analytical solution would produce the same results as a data-driven version, hence we performed the PCA method mentioned in Section 3 of the main text. As shown in Figure 2, the data-driven results do agree with the analytical ones in terms of the number of invariants overall. However, the data-driven vectors do not correspond directly to each of the conservation laws we analytically derived. Instead, the data-driven eigenvectors form a linear basis for the analytically-derived conservation laws.

## Text S4. Survey of dimensionality in various chemical mechanisms

The following two paragraphs introduce each of the 35 mechanisms included in the survey, which span orders of magnitude in complexity and serve varying use cases. At the most simplified end, we include a set of 8 representative chemistry mechanisms (each containing under 100 reactions) that have parameterized, incomplete chemistry and/or limited scope representing a narrow set of species. This low complexity set includes the KPP demonstration mechanism of simplified stratospheric chemistry Small Strato (Damian et al., 2002), the Julia Photochemical Model JPM v0.2 (Sturm & Wexler, 2022) and JPM v1.2 augmented from Sturm and Silva (2024), the Superfast Chemistry mechanism (Brown-Steiner et al., 2018; Cameron-Smith et al., 2006) obtained in KPP format from Kelp et al. (2022), a formaldehyde oxidation mechanism Form85 (Vajda et al., 1985), a simplified nitrogen and sulfur pollution mechanism POLLU (Verwer, 1994), the GEOS-Chem Hg mechanism used to simulate mercury oxidation and transport (Shah et al., 2021), and the gas-phase mechanism of an early description of tropospheric chemistry resembling the state of the science at that time (Logan et al., 1981), hereafter referred to as Logan81. On the other side of the complexity spectrum, we include the edge list of the near-explicit chemical mechanism MCM v3.3.1, with 5,832 species interacting via 16,698 reactions in a much more complete representation of atmospheric chemistry (Bloss et al., 2005; Jenkin et al., 1997, 2012, 2015; Saunders et al., 2003). We further extend this version of MCM with the Peroxy Radical Autooxidation Mechanism PRAM (Roldin, 2019; Roldin et al., 2019). The coupled pair of mechanisms MCM-PRAM has 6,041 species interacting in 18,478 reactions.

In between these extreme ends of simplification and near-explicit representation span a group of 25 chemical mechanisms with moderate complexity (roughly 40-800 species and 100-2300 reactions). The design goal of these moderate-complexity mechanisms is either to broadly capture many atmospherically relevant chemical regimes while remaining small enough to fit in the grid cells of a 3D chemical transport model or provide detailed focus on a certain set of reactive species such as halogens or oxidation of biogenic volatile organic compounds. In this moderate complexity group, we include E3SM-chem used in 3D climate modeling (Tang et al., 2025), the carbon bond mechanism CBM-Z (Zaveri & Peters, 1999), the regional acid deposition model RADM2 (Stockwell et al., 1990), SAPRC-99 (Carter, 2000), and the carbon bond mechanism CB05-TUCl updated with toluene reactions (Whitten et al., 2010) and a chlorine extension (Tanaka et al., 2003). We include two versions of the Model for Ozone and Related chemical Tracers, MOZART-4 (Emmons et al., 2010) and MOZART-T1 (Emmons et al., 2020). We include the Regional Atmospheric Chemistry Mechanism RACM (Stockwell et al., 1997) which is a predecessor to two included versions of the Community Regional Atmospheric Multiphase Mechanism CRACMM2 and CRACMM3 (Pye et al., 2023; Skipper et al., 2024). Also within the RACM family, we include the PACT-1D-HALOGENSv1.1 mechanism (Ahmed et al., 2022) which is based on RACM2 (Goliff et al., 2013) with extensions for halogen chemistry. We also compare 7 versions of GEOS-Chem (Bey et al., 2001) from v12.0.0 to v14.6.3, spanning major model updates. We include 4 mechanisms with a focus on isoprene oxidation: these are the Caltech isoprene mechanism CIM (Bates & Wennberg, 2017; Wennberg et al., 2018), a configuration of CIM with variable oxidants rather than the default fixed-oxidant approach which we call CIM-var, a manually reduced mechanism Caltech Reduced Plus RCIMv5, and an algorithmically reduced mechanism AMORE-Isoprene 2.0 with 133 species and 330 reactions, created using an automated graph theoretical approach to reduce an initially extended version of CIM (Wiser et al., 2025). We include two mechanisms from the CAABA/MECCA codebase (Sander, 2024; Sander et al., 2019): MECCAv4.6.0 and the Jülich Atmospheric Mechanism JAMv2.0 (Schultz et al., 2018). Finally we include the Chemical Reactive Intermediates mechanism CRI v2.2 (Jenkin et al., 2008, 2019; Watson et al., 2008). Table S1 contains a comprehensive overview of all mechanisms, including their relevant reference sources and notes, as well as the results of the survey: amount of coproduction, number of stoichiometric invariants, and number of emergent kinetic invariants.

For all mechanisms, we assume mass action kinetics when performing the merging operations and linear algebra described in Appendix A. However, we do note that exceptions to this could be made: a reaction in KPP can be assigned to be a constant rate (not a function of reactant concentrations) or be a function of an oxidant pool such as an $RO_2$ summation from many peroxy radical species in the system not listed as explicit reactants in the reaction (as in the MCM). For larger mechanisms, symbolic nullspaces can become extremely costly to compute. To calculate dimensionality of these mechanisms, we instead run numerical computations where rate constants are assigned numerical values for each mechanism, using 10 independently randomized trials – we find that the numerical experiments are robust to random assignment and always give identical values every time. For mechanisms like the MCM with a huge number of reactions, we make additional adjustments to the dimensional computation algorithm for the sake of computational feasibility. MCM's reaction (column) space is much larger than its species (row) space, so for each experiment iteration, we perform "randomized sketching". We project the stoichiometric matrix onto a full-rank, random Gaussian square matrix of the same dimension as the species space. By collapsing the column space, we can perform the numerical rank calculations for very large mechanisms like MCM within reasonable timeframes and without sacrificing validity. See Table S1 and Figure S2 for all mechanisms in the survey.

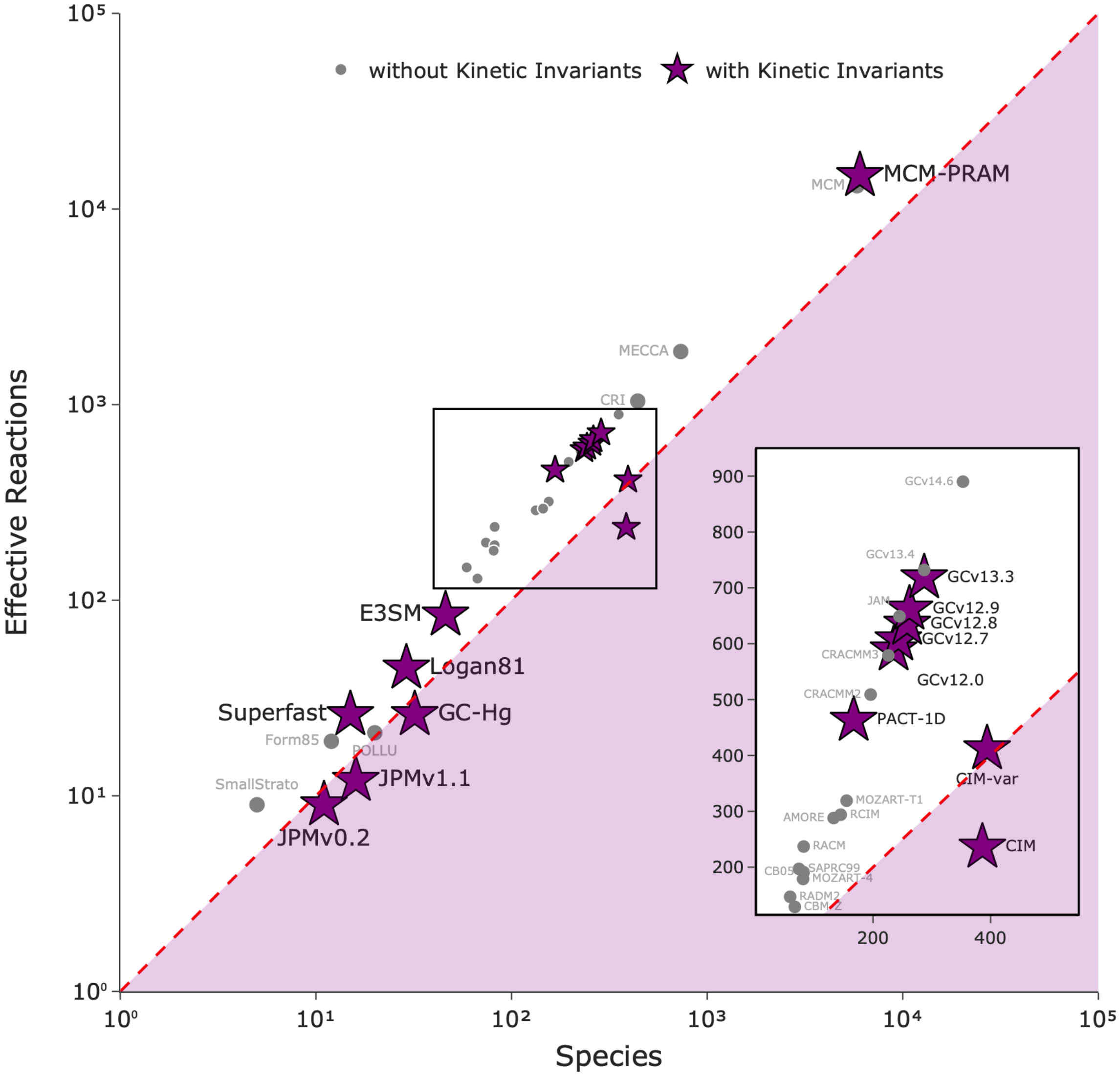

**Figure S4.** Log-scale version of Figure 3, containing all 35 mechanisms in the survey

**Table S2.** Survey of emergent conservation in 35 chemical mechanisms

| Mechanism | Short Name | Species | Reactions | Stoichiometric Invariants | Largest Merged Group | # Merged Groups | # Reactions lost to Merge | Kinetic Invariants | Broken Null Cycles | References and Source Notes |
|---|---|---|---|---|---|---|---|---|---|---|
| Small Strato | SmallStrato | 5 | 10 | 1 | 2 | 1 | 1 | 0 | 1 | (Damian et al., 2002) |
| Julia Photochem v0.2 | JPMv0.2 | 11 | 10 | 2 | 2 | 1 | 1 | 1 | 0 | (Sturm & Wexler, 2022) |
| Julia Photochem v1.2 | JPMv1.2 | 16 | 15 | 4 | 2 | 1 | 1 | 1 | 0 | Adapted from (Sturm & Silva, 2024) |
| Form85 | Form85 | 12 | 25 | 1 | 3 | 5 | 6 | 0 | 6 | (Vajda et al., 1985) is the first obtainable reference |
| Superfast | Superfast | 15 | 32 | 0 | 3 | 5 | 6 | 1 | 5 | Original reference (Cameron-Smith et al., 2006), obtained in KPP format from (Kelp et al., 2022) and added sulfate chem back based on table S2 of the supplement of (Brown-Steiner et al., 2018) |
| Verwer1994 (POLLU) | POLLU | 20 | 25 | 3 | 2 | 4 | 4 | 0 | 4 | (Verwer, 1994) is the first obtainable reference |
| Logan1981 | Logan81 | 29 | 54 | 3 | 3 | 8 | 9 | 1 | 8 | (Logan et al., 1981) table 1, gas-phase mechanism reactions 1-49. Note that we expand branching reaction and reversible reactions as their own reactions, leading to a higher count of overall reactions. The original paper used an O1D quantum yield of unity (100%) from O3 photolysis in reaction 40, which we preserve. |
| GEOS-Chem Hg | GC-Hg | 32 | 91 | 2 | 11 | 24 | 65 | 4 | 61 | (Shah et al., 2021), exact KPP version downloaded as part of GC14.6.3: https://doi.org/10.5281/zenodo.16538682 ; 3 null reactions not included in reaction count. |
| E3SM-chem | E3SM | 46 | 103 | 1 | 3 | 15 | 19 | 1 | 18 | (Tang et al., 2025) obtained from data availability statement https://github.com/E3SM-Project/E3SM/blob/master/components/eam/src/chemistry/pp_chemuci_linozv3_mam5_vbs/chem_mech.in, last access February 12, 2026 ; 1 null reaction not included in reaction count. |
| CBM-Z | CBM-Z | 67 | 139 | 7 | 3 | 9 | 10 | 0 | 10 | (Zaveri & Peters, 1999b),* ; 3 null reactions not included in reaction count. |
| RADM2 | RADM2 | 59 | 156 | 2 | 2 | 9 | 9 | 0 | 9 | (Stockwell et al., 1990),* |
| SAPRC99 | SAPRC99 | 74 | 210 | 1 | 2 | 13 | 13 | 0 | 13 | (Carter, 2000), part of KPP software examples (Damian et al., 2002; Lin et al., 2023) ; 1 null reaction not included in reaction count. |
| CB05-TUCl | CB05 | 82 | 205 | 7 | 4 | 12 | 14 | 0 | 14 | (Whitten et al., 2010),* |
| RACM | RACM | 82 | 250 | 1 | 2 | 13 | 13 | 0 | 13 | (Stockwell et al., 1997) |
| CRACMM2 | CRACMM2 | 196 | 531 | 4 | 4 | 16 | 22 | 0 | 22 | (Pye et al., 2023; Skipper et al., 2024) |
| CRACMM3 | CRACMM3 | 226 | 614 | 5 | 4 | 23 | 35 | 0 | 35 | Downloaded from the CRACMM Repo Update Oct 2025: https://doi.org/10.23719/d-tdzv. Note: adding the eliminated element tracking species adds 3 species, but only 1 additional stoichiometric invariant, indicating that 2 of 3 of the balance species do not close the atom balance |
| MOZART-4 | MOZART-4 | 81 | 196 | 0 | 3 | 16 | 17 | 0 | 17 | (Emmons et al., 2010),* |
| MOZART-T1 | MOZART-T1 | 155 | 360 | 0 | 3 | 39 | 41 | 0 | 41 | (Emmons et al., 2020),* |
| AMORE-Isoprene v2.0, 109 isoprene-derived species | AMORE | 133 | 330 | 0 | 4 | 32 | 42 | 0 | 42 | (Wiser et al., 2025) downloaded from https://doi.org/10.5281/zenodo.15026991. (Wiser et al., 2025) found that the 133-species AMORE-Isoprene mechanism had better fidelity compared to CIM than its comparable size RCIM (pink vs orange points in Figure 2 and Figure 3b of Wiser et al. (2025) – note the number of species discrepancy is because the number in the reference includes only isoprene-derived species). As neither AMORE-Isoprene nor RCIM has stoichiometric invariants or emergent conservation from kinetic invariants, neither has an effective dimensionality lower than its respective number of species. Reduced intrinsic dimensionality from emergent conservation can be ruled out as a structural cause of the difference in fidelity between RCIM and its comparatively sized AMORE-Isoprene mechanism. |
| PACT-1D-HALOGENS-v1.1 | PACT-1D | 167 | 512 | 9 | 4 | 36 | 49 | 1 | 48 | (Ahmed et al., 2022) and Zenodo https://doi.org/10.5281/zenodo.6045999 ; 1 null reaction not included in reaction count. |
| Caltech reduced plus v5 | RCIM | 145 | 379 | 0 | 5 | 64 | 85 | 0 | 85 | (Wennberg et al., 2018) downloaded from (Bates & Wennberg, 2017) |
| Caltech Isoprene Mechanism Full v5 | CIM | 386 | 886 | 0 | 15 | 195 | 650 | 150 | 500 | (Wennberg et al., 2018) downloaded from (Bates & Wennberg, 2017) |
| Caltech Isoprene Mechanism, Variable Oxidants | CIM-var | 394 | 886 | 2 | 8 | 284 | 474 | 17 | 457 | (Wennberg et al., 2018) downloaded from (Bates & Wennberg, 2017) |
| Jülich Atmospheric Mechanism JAM002b | JAM | 245 | 702 | 4 | 4 | 41 | 53 | 0 | 53 | (Schultz et al., 2018) downloaded in the supplement of (Sander, 2024) |
| GEOS-Chem v12.0.0 | GCv12.0 | 235 | 725 | 6 | 6 | 101 | 136 | 1 | 135 | Standard mechanism downloaded from https://doi.org/10.5281/zenodo.1343547 |
| GEOS-Chem v12.7.0 | GCv12.7 | 243 | 750 | 6 | 6 | 108 | 143 | 1 | 142 | Standard mechanism downloaded from https://doi.org/10.5281/zenodo.3634864 |
| GEOS-Chem v12.8.0 | GCv12.8 | 258 | 825 | 6 | 6 | 138 | 190 | 1 | 189 | Standard mechanism downloaded from https://doi.org/10.5281/zenodo.3784796 |
| GEOS-Chem v12.9.0 | GCv12.9 | 262 | 850 | 8 | 6 | 138 | 188 | 2 | 186 | Standard mechanism downloaded from https://doi.org/10.5281/zenodo.3950327. Ablation study with no P/L species showed the mechanism still has 2 kinetic invariants |
| GEOS-Chem v13.3.0 | GCv13.3 | 287 | 903 | 10 | 6 | 140 | 185 | 2 | 183 | Full chem mechanism downloaded from https://doi.org/10.5281/zenodo.7254231 |
| GEOS-Chem v13.4.0 | GCv13.4 | 287 | 913 | 10 | 6 | 138 | 181 | 0 | 181 | Full chem mechanism downloaded from https://doi.org/10.5281/zenodo.7254268 |
| GEOS-Chem v14.6.3 | GCv14.6 | 353 | 1058 | 9 | 5 | 123 | 168 | 0 | 168 | Full chem mechanism downloaded from https://doi.org/10.5281/zenodo.16538682 |
| CRI v2.2 | CRI | 442 | 1258 | 2 | 6 | 179 | 216 | 0 | 216 | (Jenkin et al., 2008, 2019; Watson et al., 2008) downloaded from https://www.mcm.york.ac.uk/CRI last access November 6, 2025. |
| CAABA /MECCA v4.6.0 | MECCA | 733 | 2323 | 9 | 8 | 361 | 456 | 0 | 456 | Downloaded as gas.kpp mechanism in the supplement of (Sander, 2024). The most recent code is available on gitlab: https://gitlab.com/RolfSander/caaba-mecca |
| MCM v3.3.1 | MCM | 5832 | 16698 | 1 | 7 | 2617 | 3558 | 0 | 3558 | (Bloss et al., 2005; Jenkin et al., 1997, 2012, 2015; Saunders et al., 2003) downloaded from https://www.mcm.york.ac.uk/MCM/ last access November 6, 2025. |
| MCM with PRAM | MCM-PRAM | 6041 | 18478 | 1 | 7 | 2654 | 3671 | 3 | 3668 | MCM v3.3.1 as above, paired with PRAM downloaded in KPP format (Roldin, 2019; Roldin et al., 2019) |

*downloaded in KPP format from the BOXMOX (Knote et al., 2015) repository: https://mbees.med.uni-augsburg.de/gitlab/mbees/boxmox/-/tree/master/models?ref_type=heads, last access January 6, 2025

Note: the stoichiometric invariant column counts invariants present in the cokernel of the stoichiometric matrix (all structural invariants), which in mechanisms that write branching as pre-merged reactions, could also include invariants of kinetic origin. The kinetic invariant column counts only emergent invariants not detectable from mechanism static structure.